\documentclass[aps,pra,showpacs,preprint,floatfix,12pt]{revtex4-1}

\usepackage{mathrsfs}
\usepackage{graphicx}
\usepackage{epstopdf}
\usepackage{amsmath}
\usepackage{latexsym}
\usepackage{amsfonts}
\usepackage{amssymb}
\usepackage{array}
\usepackage{subfigure}
\usepackage{braket}

\newcommand{\beq}{\begin{equation}}
\newcommand{\eeq}{\end{equation}}
\newcommand{\ba}{\begin{array}}
\newcommand{\ea}{\end{array}}
\newcommand{\bea}{\begin{eqnarray}}
\newcommand{\eea}{\end{eqnarray}}

\begin{document}

\title[]{Comment on ``Direct photodetachment of F$^-$ by mid-infrared few-cycle
femtosecond laser pulses''}
\author{G. F. Gribakin}
\email{g.gribakin@qub.ac.uk}
\author{S. M. K. Law}
\email[]{slaw08@qub.ac.uk}
\affiliation{Center for Theoretical Atomic, Molecular and Optical Physics,
Queen's University Belfast, Belfast BT7 1NN, United Kingdom}

\begin{abstract}
Multiphoton detachment of F$^-$ by strong few-cycle laser pulses was studied by
Shearer and Monteith using a Keldysh-type approach [Phys. Rev. A \textbf{88},
033415 (2013)]. We believe that this work contained errors in the
calculation of the detachment amplitude and photoelectron spectra. We describe
the necessary corrections to the theory and show that the results, in
particular, the interference features of the photoelectron spectra, appear
noticeably different.
\end{abstract}

\maketitle

In Ref.~\cite{shearer} direct photodetachment of F$^-$ by a strong
linearly-polarized laser field was considered using the Keldysh-type approach
(KTA) \cite{gleb} generalized to few-cycle pulses \cite{shearer1}. Such
methods are useful in general for studying strong-field effects in few-cycle
pulses, see, e.g., Ref.~\cite{milosevic}. The study was performed for an
$N$-cycle pulse with the vector potential of the form
\begin{equation}\label{eq:A}
\textbf{A}(t)=A_0 \hat{\textbf{z}}\sin^2\left(\frac{\omega t}{2N}\right)
\sin(\omega t + \alpha) \, ,
\end{equation}
where $A_0$ is the peak amplitude, $\omega$ is the carrier frequency and
$\alpha$ is the carrier-envelope phase (CEP). Photoelectron momentum, angular
and energy distributions were generated for a $N=4$ cycle laser pulse with a
range of peak intensities and mid-infrared wavelengths, while examining effects
related to above-threshold channel closures and variation of the CEP.

A calculation similar to that of Ref.~\cite{shearer} was also used to identify
the effect of electron rescattering in short-pulse multiphoton detachment
from F$^-$ computed using the R-matrix with time dependence (RMT) method
\cite{ola}. Subsequently, an error in the KTA calculations was uncovered
\cite{erratum}. It concerned the phase factors of the contributions to the
detachment amplitude that arose from successive saddle points in the KTA
calculation for a $p$-wave electron. Upon correction, the KTA results showed a
better agreement with the RMT photoelectron spectra \cite{erratum,ola}.
We believe that the same error affected the results of Ref.~\cite{shearer}.
In this comment we show that the interference features of the photoelectron
momentum and angular distributions, and the energy spectra are distinctly
different from those of Ref.~\cite{shearer} when calculated correctly. We use
atomic units throughout, unless stated otherwise.

Using the Keldysh-like approach \cite{gleb} for the $N$-cycle pulse
(\ref{eq:A}), one finds the detachment amplitude for an initial state with
orbital and magnetic quantum numbers $l$ and $m$, as (see Eq.~(16) of
Ref.~\cite{shearer}),
\begin{equation}\label{eq:ampcorrect}
A_\textbf{p}=-(2\pi)^{3/2} A \sum_{\mu=1}^{2(N+1)} (\pm)^{l}
Y_{lm}(\hat{\textbf{p}}_\mu) \frac{\exp[if(t_\mu)]}{\sqrt{-if''(t_\mu)}},
\end{equation}
where \textbf{p} is the final photoelectron momentum and $A$ is the asymptotic
normalization constant of the bound electron wave function (for F$^-$ we use
$A=0.7$ \cite{nikitin}). Equation (\ref{eq:ampcorrect}) involves a sum over
$2(N+1)$ saddle points $t_\mu$ in the complex time plane, ${\bf p}_\mu$ and
$f(t_\mu)$ being the classical electron momentum and action respectively,
evaluated at the saddle points. The terms in the sum in
Eq.~(\ref{eq:ampcorrect}) contain a phase factor $(\pm)^l\equiv (\pm 1)^l$ that
alternates (for an odd $l$) between the contributions from successive saddle
points. When the spherical function
$Y_{lm}(\hat{\textbf{p}}_\mu)\equiv Y_{lm}(\Theta ,\varphi )$ in
Eq.~(\ref{eq:ampcorrect}) is evaluated for complex vectors ${\bf p}_\mu$, the
polar angle $\Theta$ is determined by
\begin{equation}\label{eq:complextheta}
\cos \Theta = \sqrt{1+p_{\perp}^{2}/\kappa_j^2},\quad
\sin \Theta = \mp i p_\perp /\kappa_j,
\end{equation}
where $p_\perp $ is the component ${\bf p}$ perpendicular to the $z$ axis, and
$\kappa_j =\sqrt{2|E_j|}$ parameterizes the energy $E_j$ of the bound state for
each fine-structure component $j=3/2,~1/2$ of F$^-$ ($l=1$). The sign in
$\sin \Theta $ alternates in the opposite way to $(\pm)$ in
Eq.~(\ref{eq:ampcorrect}) and gives rise to an additional $m$-dependent
phase factor. The final expression for the differential detachment probability
of an electron from the state $l,m$ reads
\begin{equation}\label{eq:glebshort}
\frac{dw_{lm}^{(j)}}{d^3{\bf p}}=\frac{A^2}{4\pi}(2l+1)\frac{(l-|m|)!}{(l+|m|)!}
\left|P_l^{|m|}\left(\sqrt{1+p_\perp^2/\kappa_j^2}\right) \right|^2 \left|
\sum_{\mu=1}^{2(N+1)} (\pm)^{l+m} \frac{\exp[if(t_\mu)]}{\sqrt{-if''(t_\mu)}}
\right|^2,
\end{equation}
where $P_l^{|m|}$ is the associated Legendre function.
The superscript $j$ is introduced into expression (\ref{eq:glebshort}) to
indicate the detachment from the spin-orbit sublevel $j$ of the ion,
which contributes with the statistical factor $(2j+1)/(2l+1)$ to the
total detachment probability. The numerical values of $\kappa_j$ for each
fine-structure state of F$^-$ are $\kappa_{3/2}=0.4998$ and
$\kappa_{1/2}=0.5035$ (using the electron affinity of F$^-$ from
Ref.~\cite{hotop}). Note that Eq.~(\ref{eq:glebshort}) takes a similar form to
Eq.~(33) from Ref.~\cite{gleb} in the case of the long periodic pulse.

The photoelectron momentum densities are axially symmetric and can be obtained
from Eq.~(\ref{eq:glebshort}) by taking ${\bf p}$ in the Cartesian
momentum plane ($p_x,p_z$),
\begin{equation}\label{eq:momeq}
\sum_j\frac{2j+1}{2l+1}\sum _{m=-l}^{l}\frac{dw_{lm}^{(j)}}{d^3{\bf p}}.
\end{equation}
The photoelectron angular distribution is obtained by integrating
Eq.~(\ref{eq:glebshort}) over the electron energy $E_e=p^2/2$,
\begin{equation}\label{eq:angdist}
\frac{dw}{d\theta}=2\pi\sin\theta \sum_j \sum_{m=-l}^{l} \frac{2j+1}{2l+1}
\int_0^{\infty} \frac{dw_{lm}^{(j)}}{d^3{\bf p}} dE_e,
\end{equation}
where $\theta $ is the polar angle. The photoelectron energy spectrum is
given by
\begin{equation}\label{eq:energyspec}
\frac{dw}{dE_e}=2\pi \sum_j \sum_{m=-l}^{l} \frac{2j+1}{2l+1}
\int_0^{\pi} \frac{dw_{lm}^{(j)}}{d^3 \textbf{p}} \sin\theta d\theta,
\end{equation}
and the total detachment probability is
\begin{equation}\label{eq:totalprobeq}
w=\int_0^{\infty} \frac{dw}{dE_e} dE_e\equiv
\int_0^{\pi} \frac{dw}{d\theta} d\theta .
\end{equation}

In Ref.~\cite{ola} and, we believe, in Ref.~\cite{shearer} too, the presence of
the phase factor $(\pm)^{l+m}$ in the sum over the saddle points in
Eq.~(\ref{eq:glebshort}) was overlooked in the calculations. As a result, the
detachment probability for $p$ electrons
($l=1$) was computed correctly for $m=\pm 1$, but incorrectly for $m=0$, with
the interference contributions between the odd and even saddle points
added with the wrong sign. Since $m=0$ electron states give a dominant
contribution to the detachment signal, this error affected the interference
patterns of the photoelectron momentum and energy distributions presented in
Ref.~\cite{shearer} (see \cite{ola} and erratum \cite{erratum}).
In addition, we have found that the magnitudes of the photoelectron angular and
energy spectra in Figs. 4--7 of Ref.~\cite{shearer} are incorrect. This is in
part due to the extra spin factor 2 in Eq. (19) of Ref.~\cite{shearer}, which
was erroneously retained when accounting for the fine-structure splitting
in Eq. (23) of Ref.~\cite{shearer}, and also affected the KTA results in
Ref.~\cite{ola}.
The purpose of this Comment is to present correct photoelectron distributions
for the same wavelengths and other laser-pulse parameters as used originally in
Ref.~\cite{shearer}.

\begin{figure}[ht]
\hspace*{-0.5in}
\mbox{\subfigure{\includegraphics[width=2.45in]{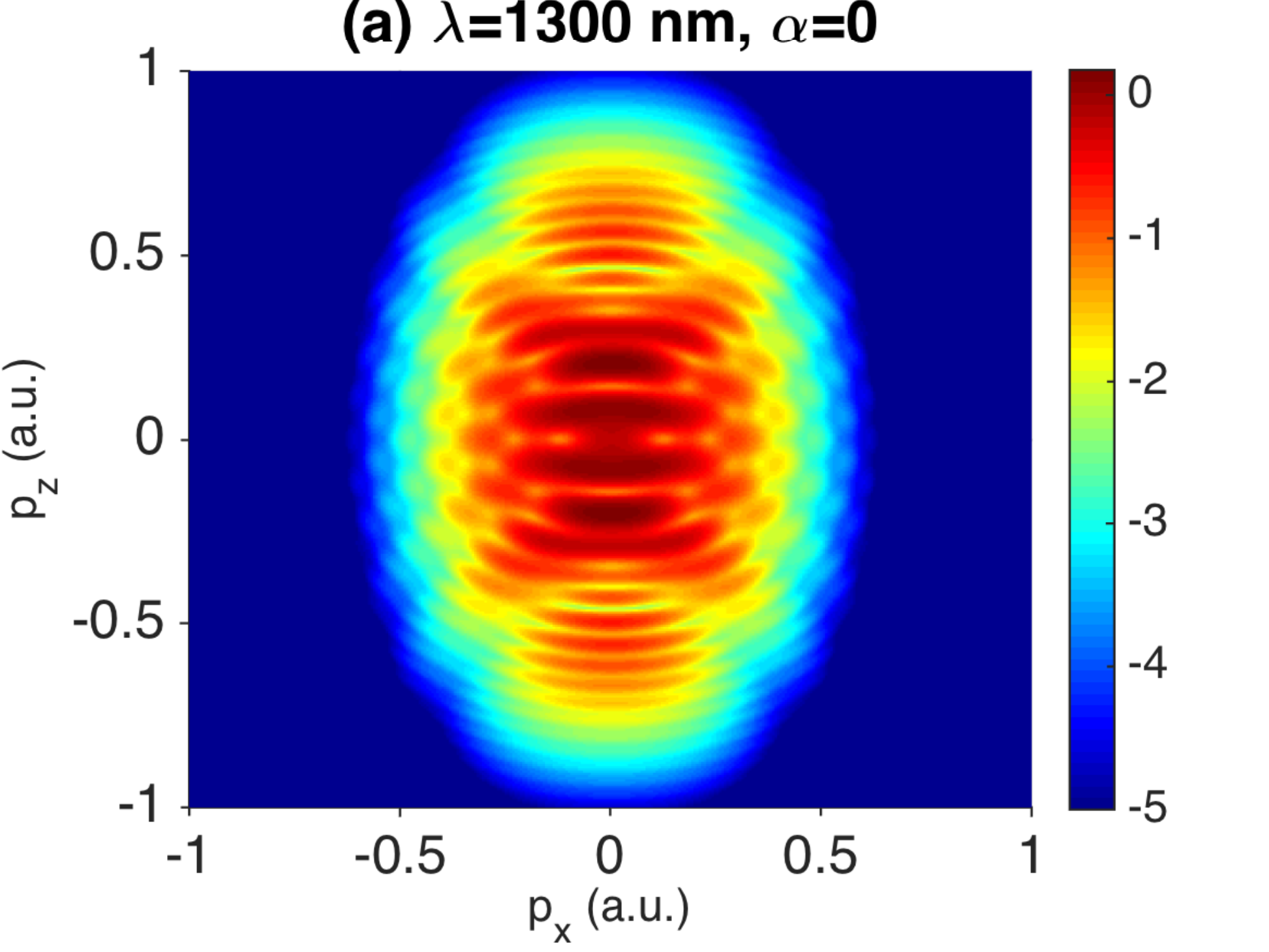}}
\subfigure{\includegraphics[width=2.45in]{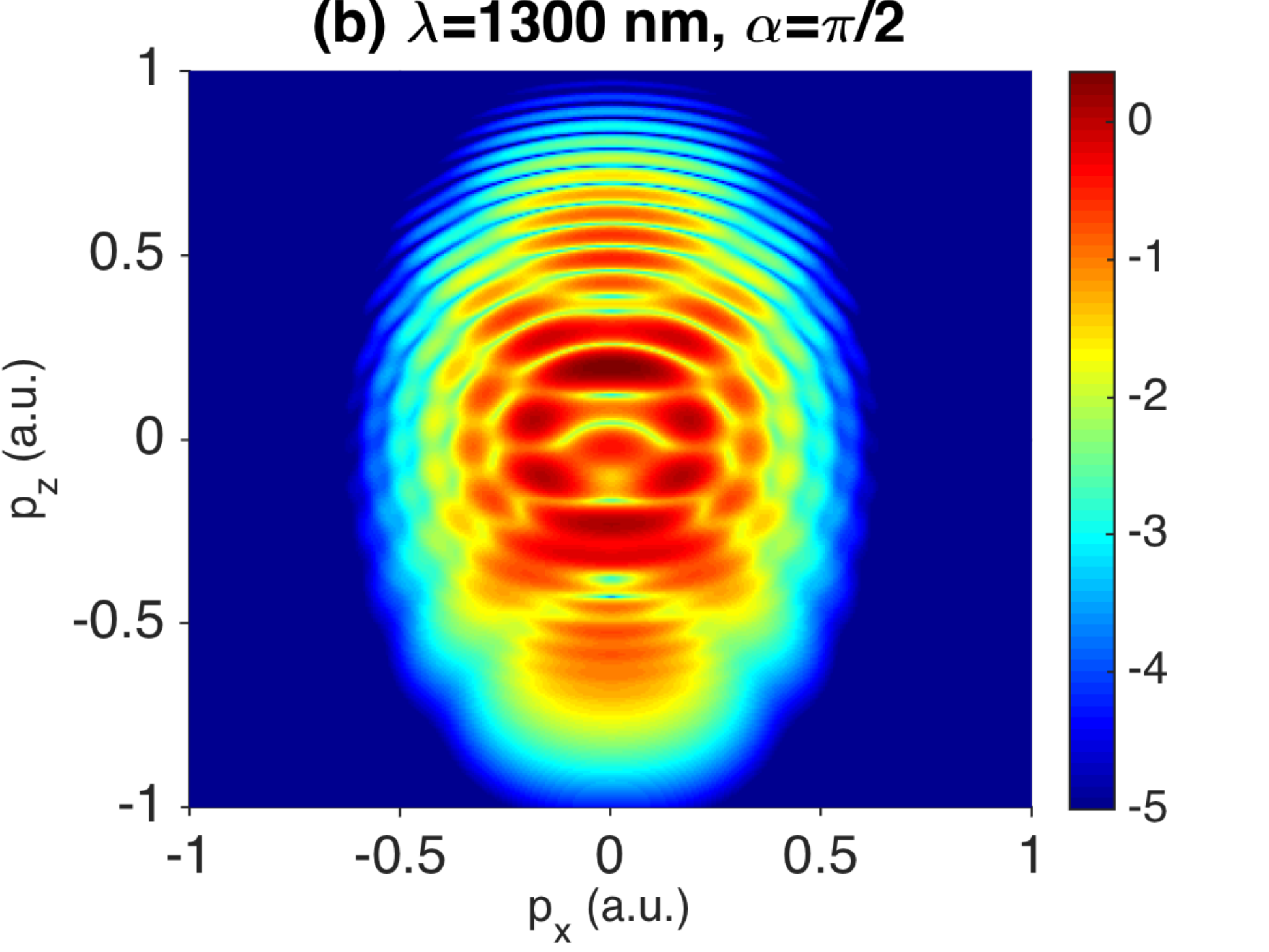}}
\subfigure{\includegraphics[width=2.45in]{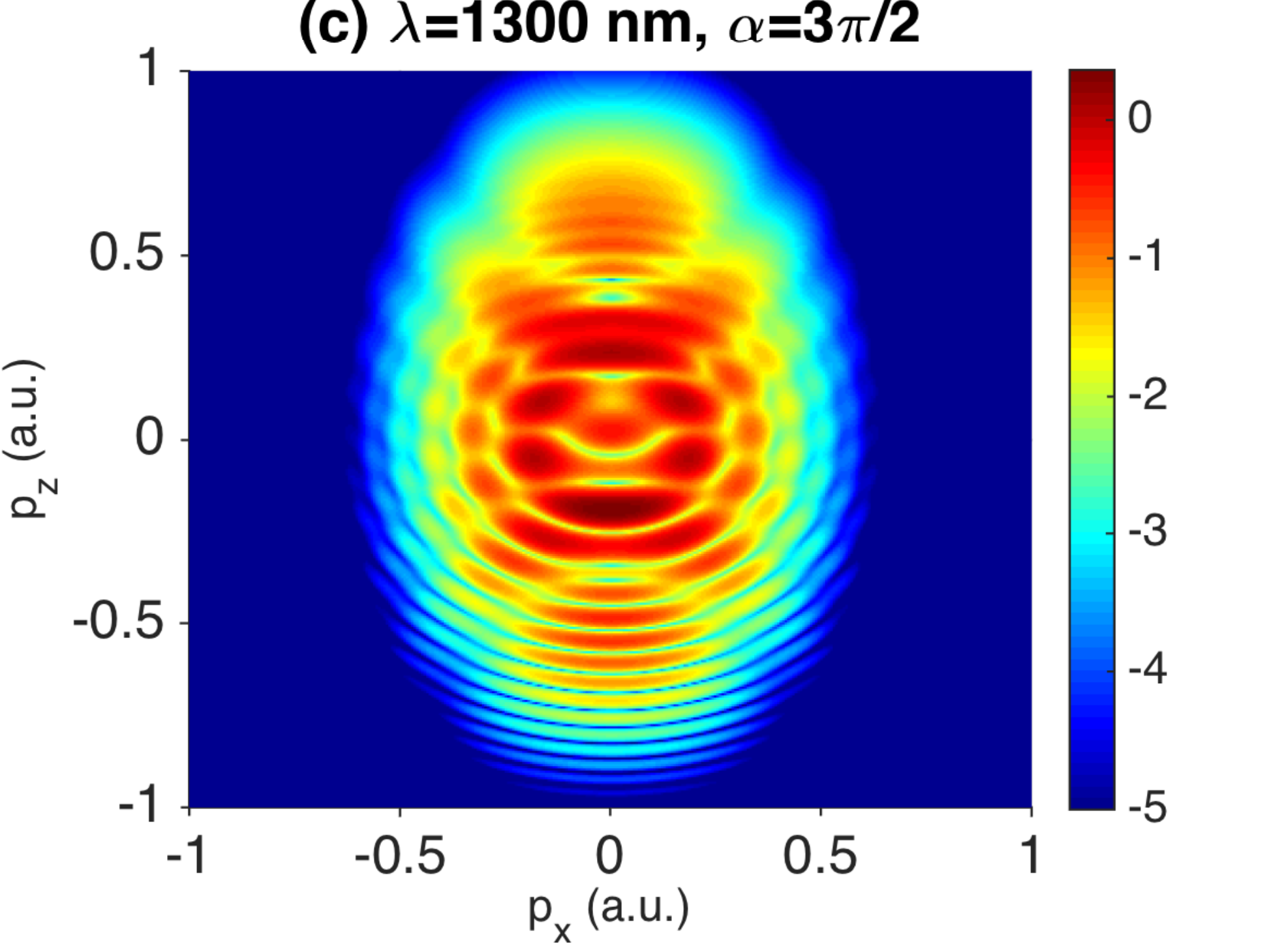}}}
\hspace*{-0.5in}
\mbox{\subfigure{\includegraphics[width=2.45in]{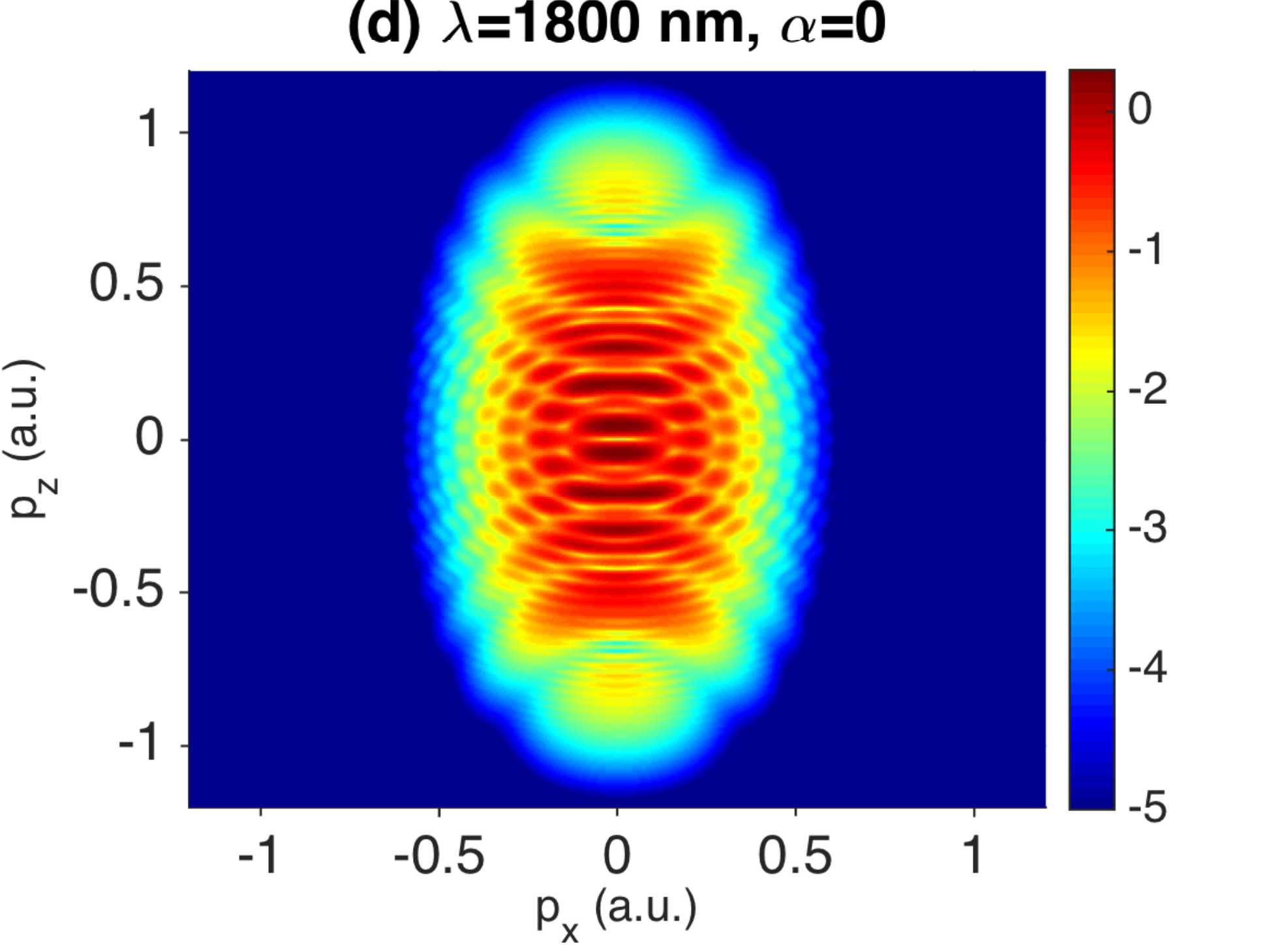}}
\subfigure{\includegraphics[width=2.45in]{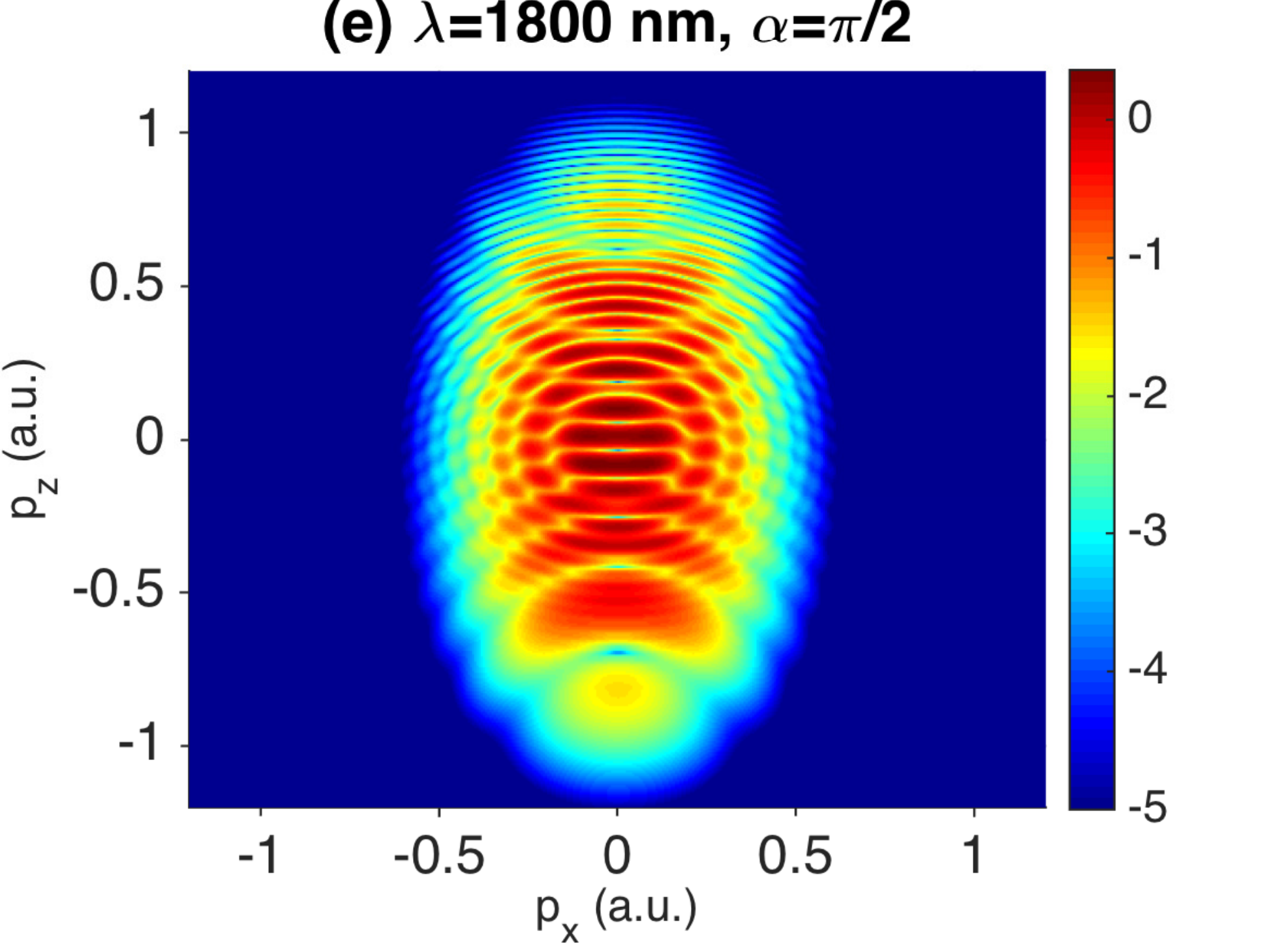}}
\subfigure{\includegraphics[width=2.45in]{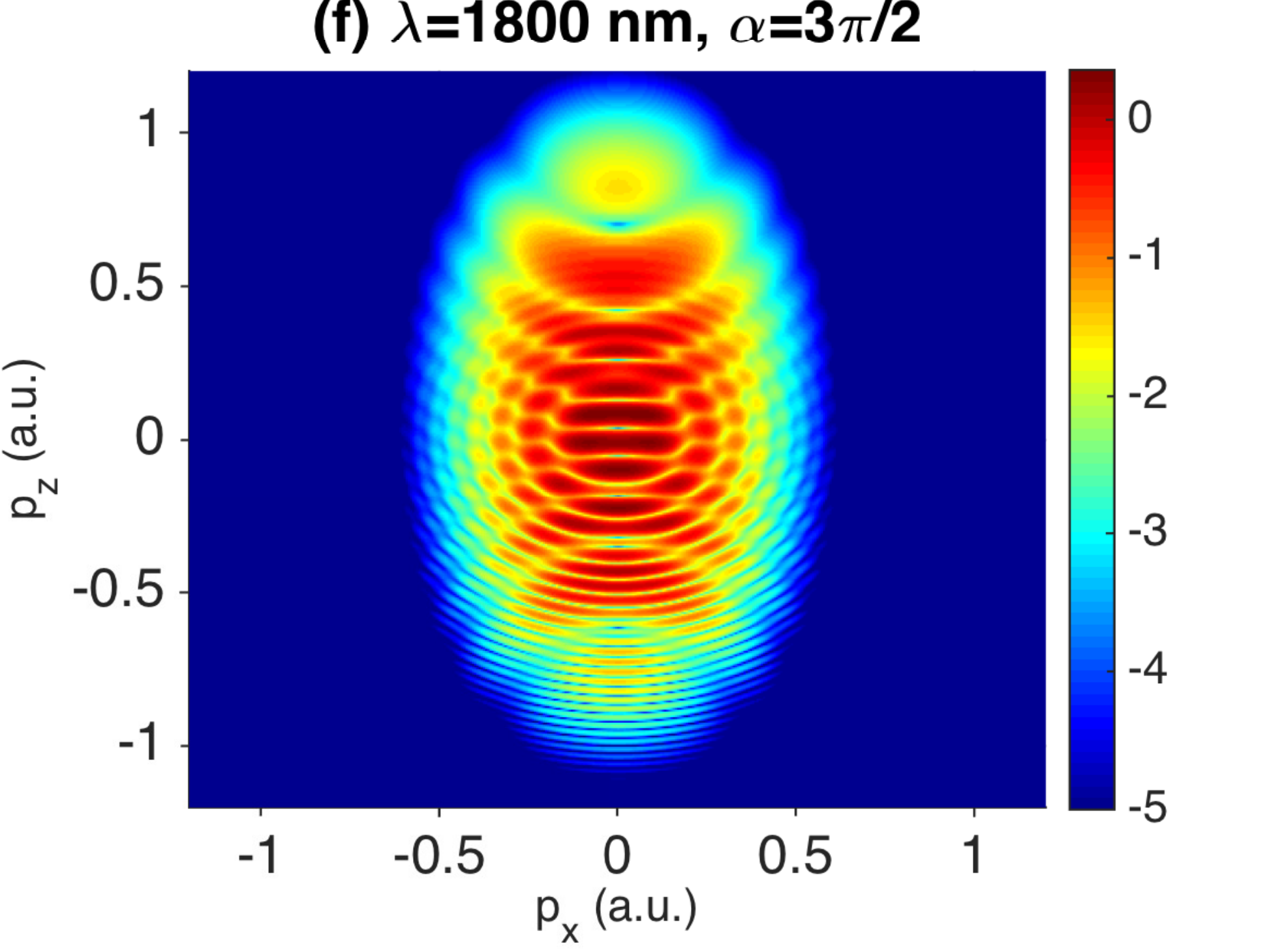}}}
\caption{(Color online) Logarithmic momentum densities for photodetachment of
F$^-$ by a four-cycle laser pulse with peak intensity of
$1.3\times 10^{13}$~W/cm$^2$. The top and bottom rows corresponds to
$\lambda = 1300$ and 1800~nm, respectively, calculated for $\alpha =0$ [(a) and
(d)], $\alpha =\pi /2$ [(b) and (e)] and $\alpha=3\pi/2$ [(c) and (f)].}
\label{fig:momentum}
\end{figure}

Figure~\ref{fig:momentum} displays logarithmic photoelectron momentum densities
for photodetachment of F$^-$ by a four-cycle pulse with peak intensity of
$1.3 \times 10^{13}$~W/cm$^2$ and carrier wavelength of 1300 and 1800~nm,
for CEP values $\alpha=0$, $\pi/2$, and $3\pi/2$. Compared with Fig.~2
of Ref.~\cite{shearer}, the correct interference patterns appear more diffuse,
lacking any sharp features.
Figures~\ref{fig:momentum} (a) and (d) show closer agreement with the momentum
densities predicted by the RMT method \cite{ola}.
At the same time, the overall forward-backward asymmetry along the $p_z$
direction (for $\alpha =\pi/2$ and $3\pi/2$) is generally unaffected, since
this characteristic depends on the symmetry of the laser field only.

\begin{figure}[ht]
\hspace*{-0.5in}
\mbox{\subfigure{\includegraphics[width=2.45in]{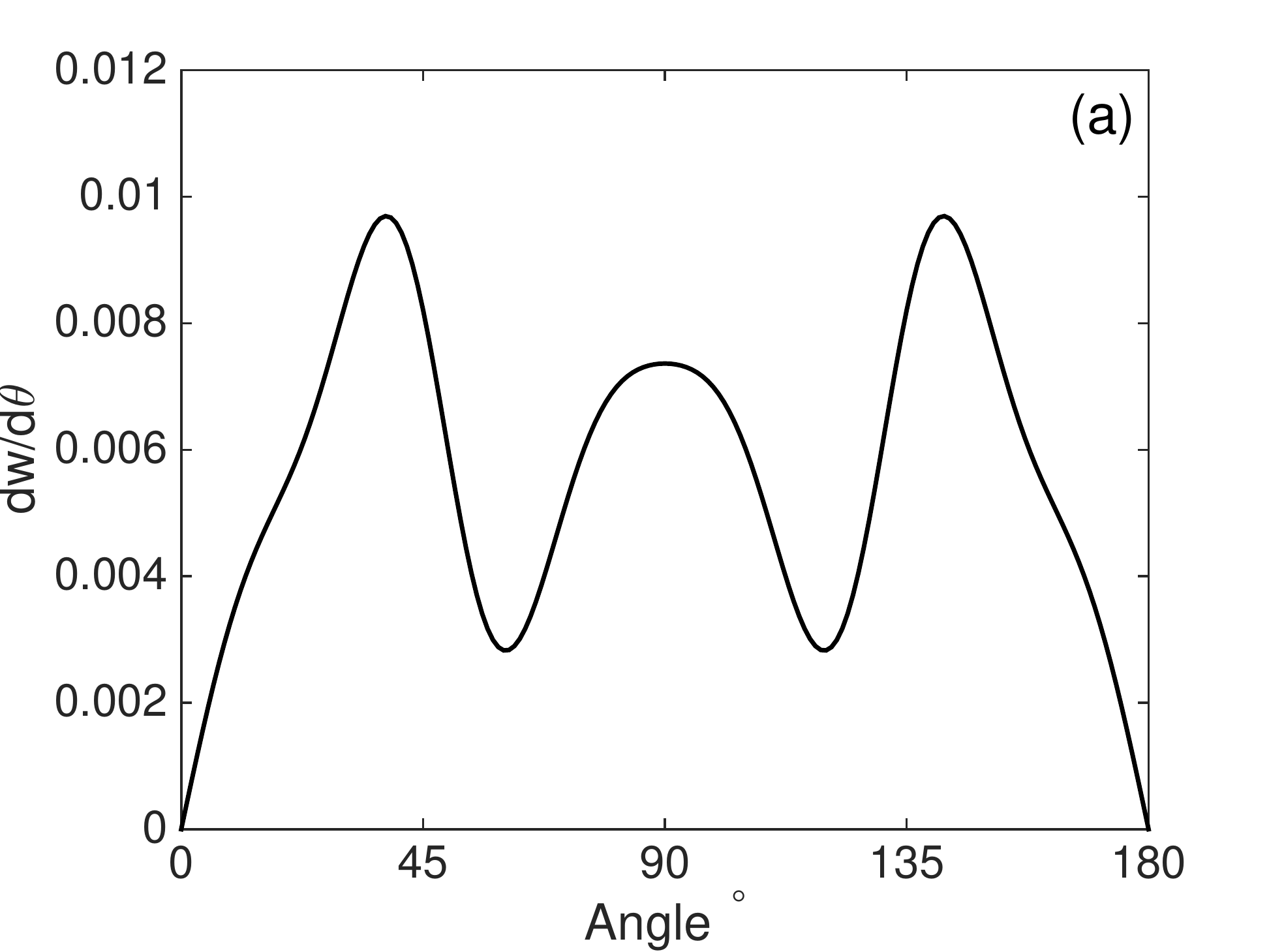}}
\subfigure{\includegraphics[width=2.45in]{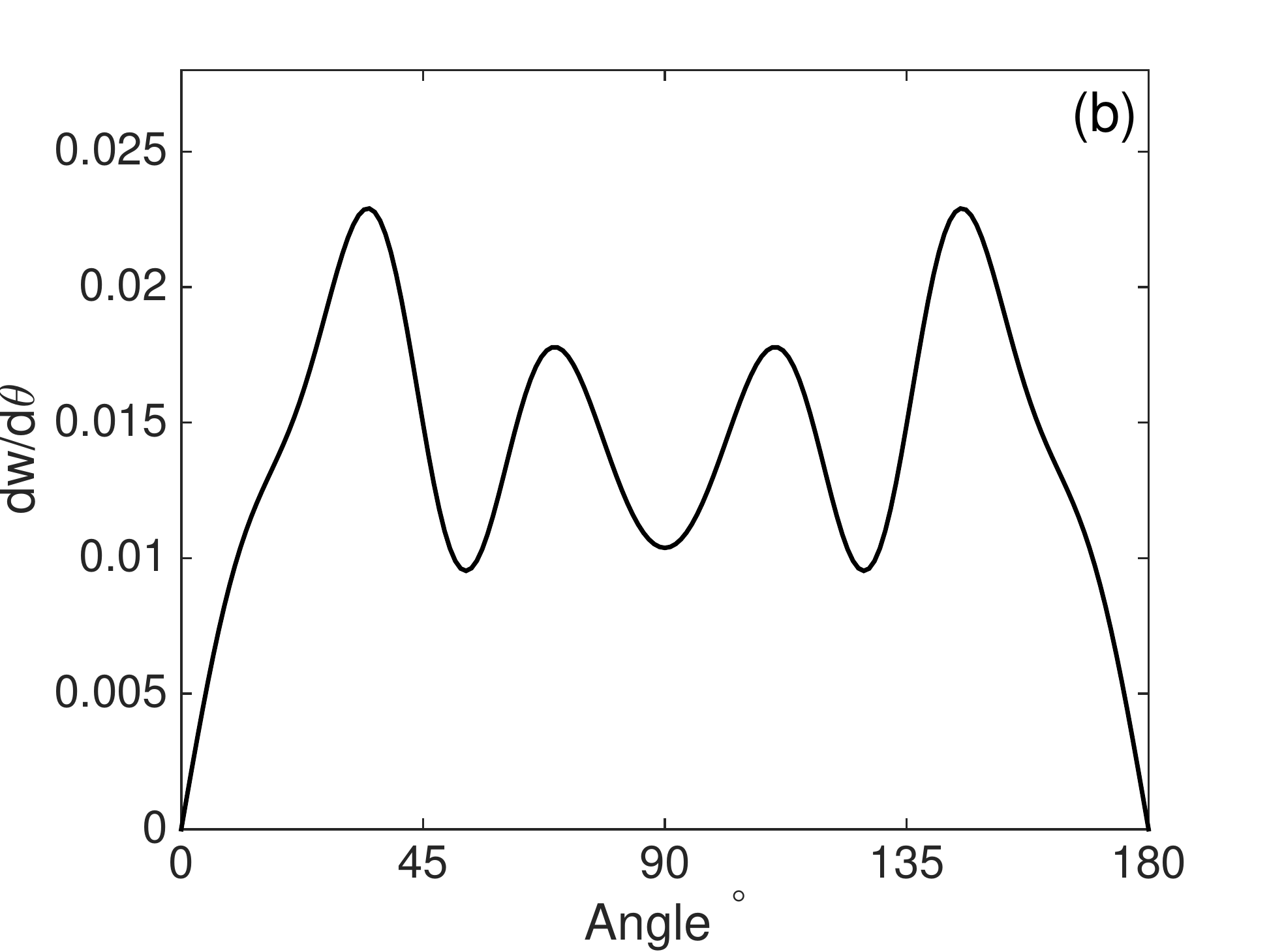}}
\subfigure{\includegraphics[width=2.45in]{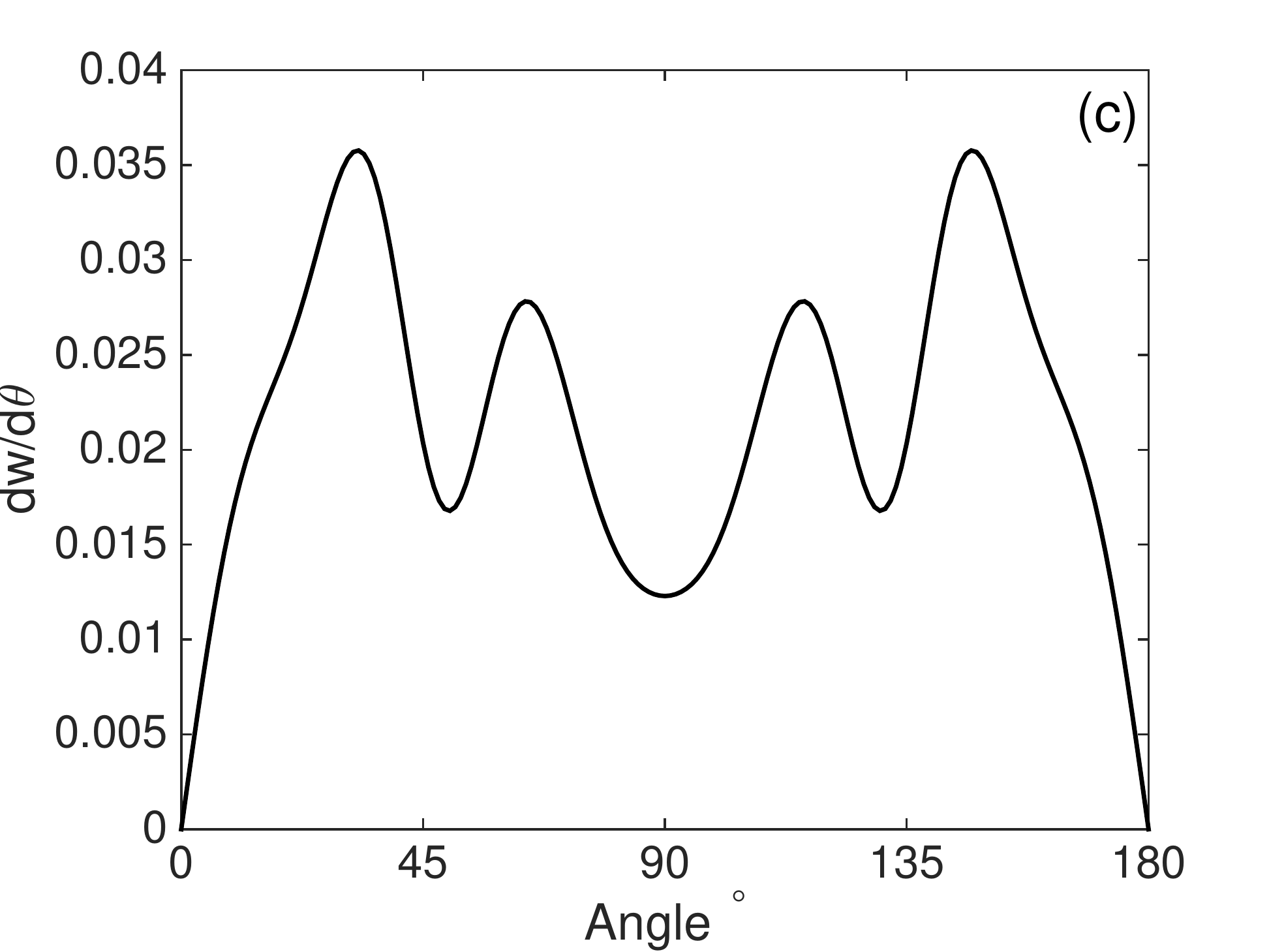}}}
\hspace*{-0.5in}
\mbox{\subfigure{\includegraphics[width=2.45in]{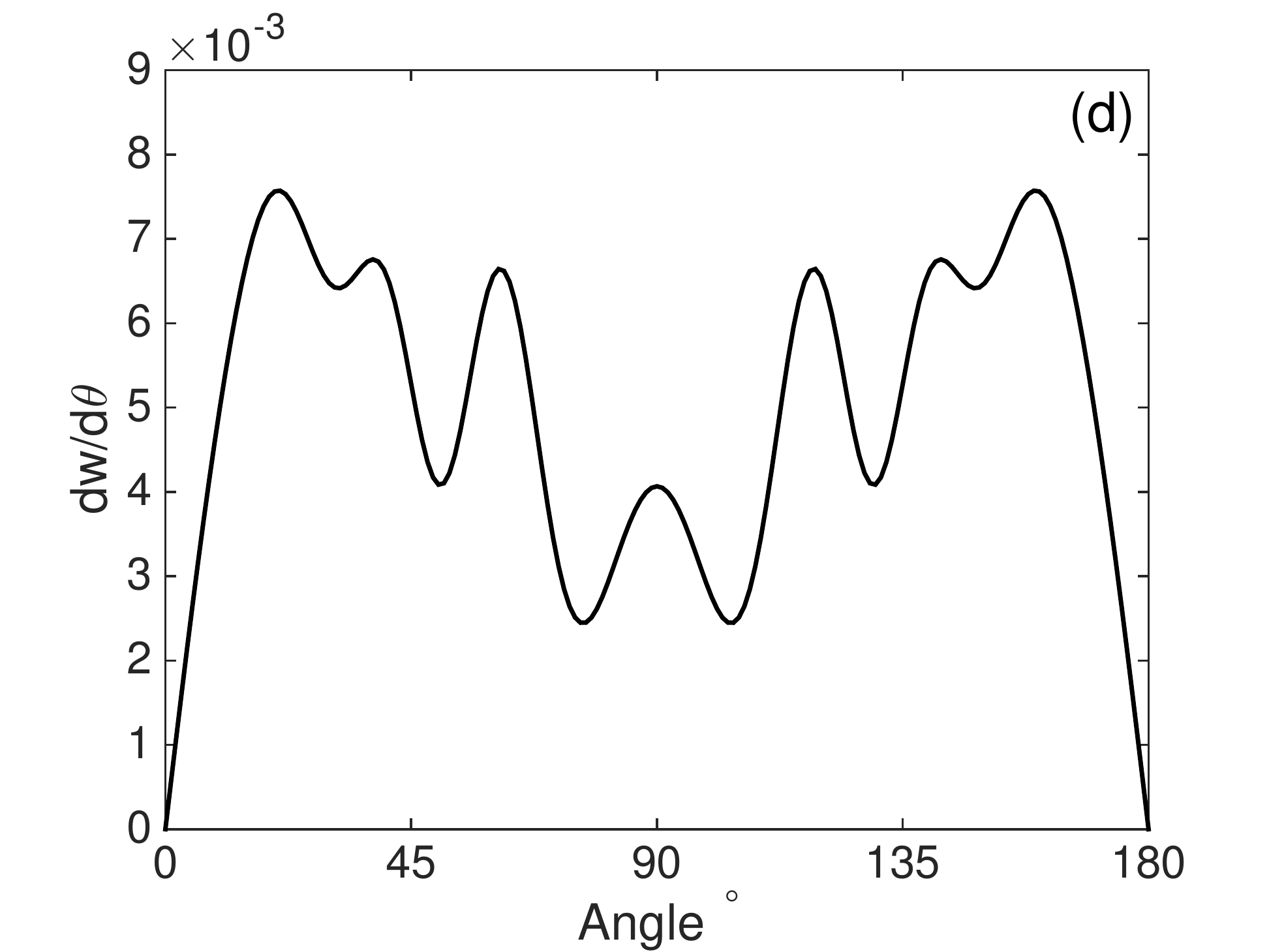}}
\subfigure{\includegraphics[width=2.45in]{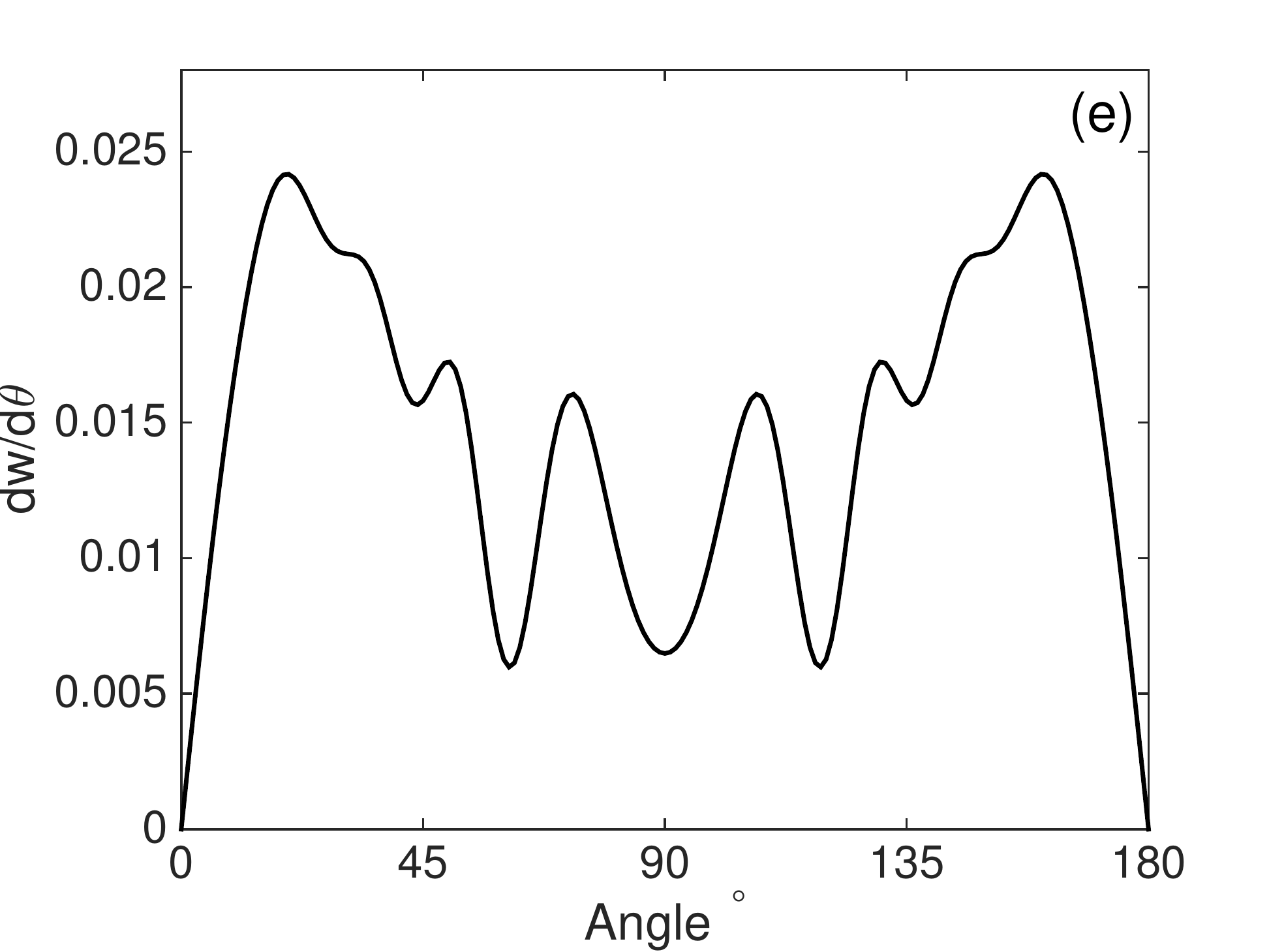}}
\subfigure{\includegraphics[width=2.45in]{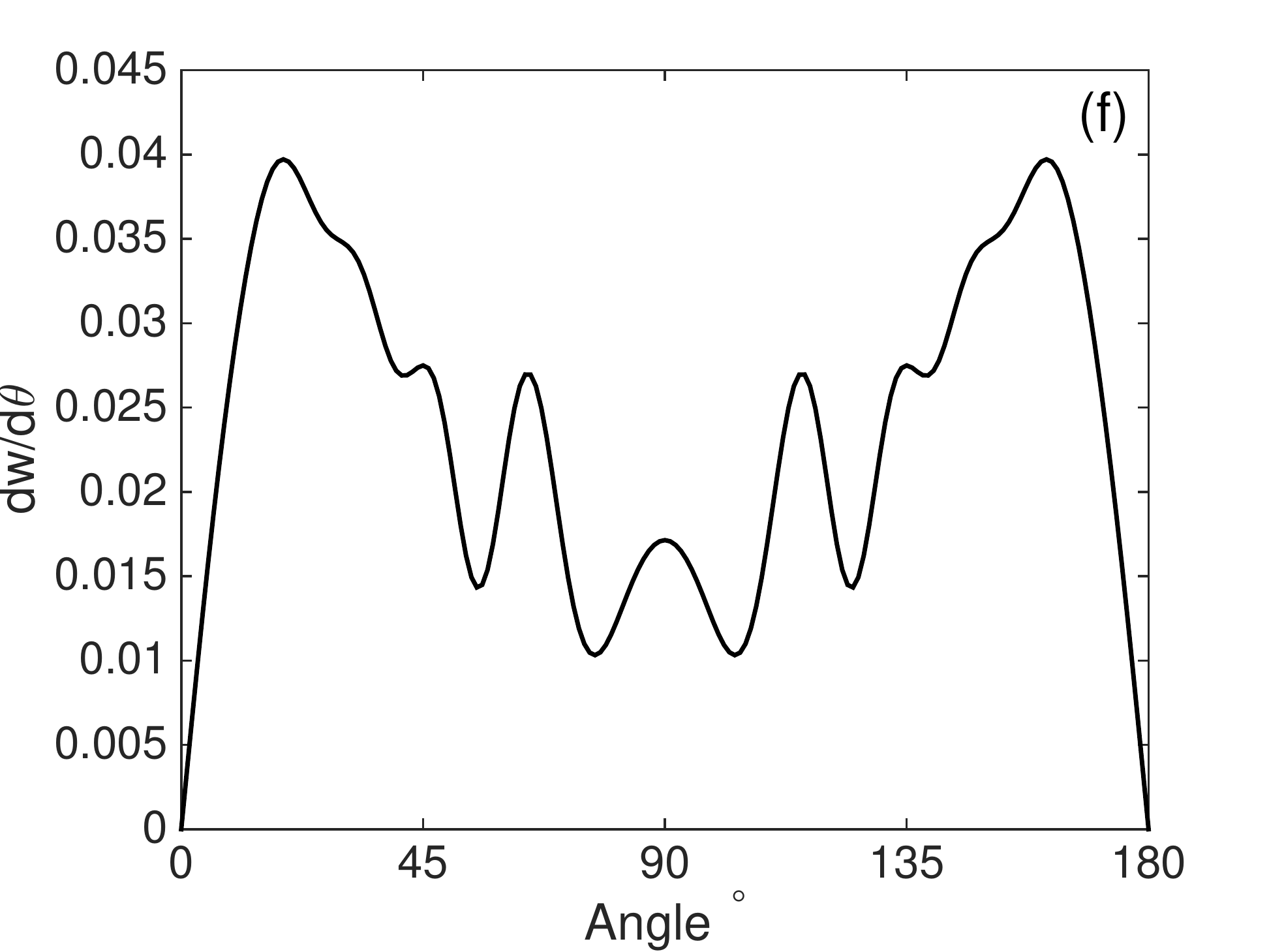}}}
\caption{Photoelectron angular distributions for F$^-$ for a four-cycle pulse
with $\lambda = 1300$ (top row) and 1800 nm (bottom row), CEP $\alpha$=0, and
peak intensities $7.7\times10^{12}$, $1.1\times 10^{13}$, and
$1.3\times 10^{13}$~W/cm$^2$ (left, central, and right columns respectively);
(a)--(c) correspond to $n_{\rm min}=5$, 6, and 6, and (d)--(f) to
$n_{\rm min} =9$, 10, and 11-photon detachment, respectively.}
\label{fig:angle}
\end{figure}

Figure~\ref{fig:angle} shows photoelectron angular distributions for F$^-$
for a four-cycle pulse with CEP $\alpha =0$, wavelengths 1300 and 1800~nm and
intensities of $7.7\times 10^{12}$, $1.1\times 10^{13}$ and
$1.3\times 10^{13}$~W/cm$^2$. Because of the errors mentioned earlier,
these angular distributions are very different, both in shape and magnitude,
from the (incorrect) results in Fig. 5 of Ref.~\cite{shearer}. The oscillatory
structure of the distributions is related to the minimum number of photons
 that needs to be absorbed near the peak of the pulse, $n_{\rm min}$
(determined by the integer part of $(U_{\bf p} + |E_j|)/\omega +1$ for a given
ponderomotive energy $U_{\bf p}=A_0^2/4$). Analysis of
Fig.~\ref{fig:angle} shows that the angular distributions are characterized by
a local maximum (minimum) in the direction perpendicular to the field
($\theta = \pi/2$) when $n_{\rm min}$ is odd (even). This can be seen in
Fig.~\ref{fig:angle} (a), (d) and (f) corresponding to $n_{\rm min}=5$, 9 and
11, respectively, and in contrast to the original (incorrect) results of
\cite{shearer} where a central minimum for odd $n_{\rm min}$ was noted.
The effect of channel closure with increasing intensity gives rise to
even $n_{\rm min}$=6, 6 and 10 and a minimum at $\theta =\pi /2$, as seen in
Fig.~\ref{fig:angle} (b), (c), and (e), respectively. This behaviour is in
agreement with the observation that for a long periodic pulse the $n$-photon
detachment rate is exactly zero at $\theta=\pi/2$ for odd $n+l+m$ \cite{gleb},
and the fact that $m=0$ dominates the photoelectron spectrum.
Figure~\ref{fig:angle} also indicates that electron emission at angles close to
the direction of the field (i.e., within $0\leq\theta\leq 45^{\circ}$ and
$135^{\circ}\leq\theta\leq 180^{\circ}$) is much stronger here in comparison to
Ref.~\cite{shearer}, and in better accord with the momentum maps in
Fig.~\ref{fig:momentum}.

\begin{figure}[ht]
\hspace*{-0.5in}
\mbox{\subfigure{\includegraphics[width=2.45in]{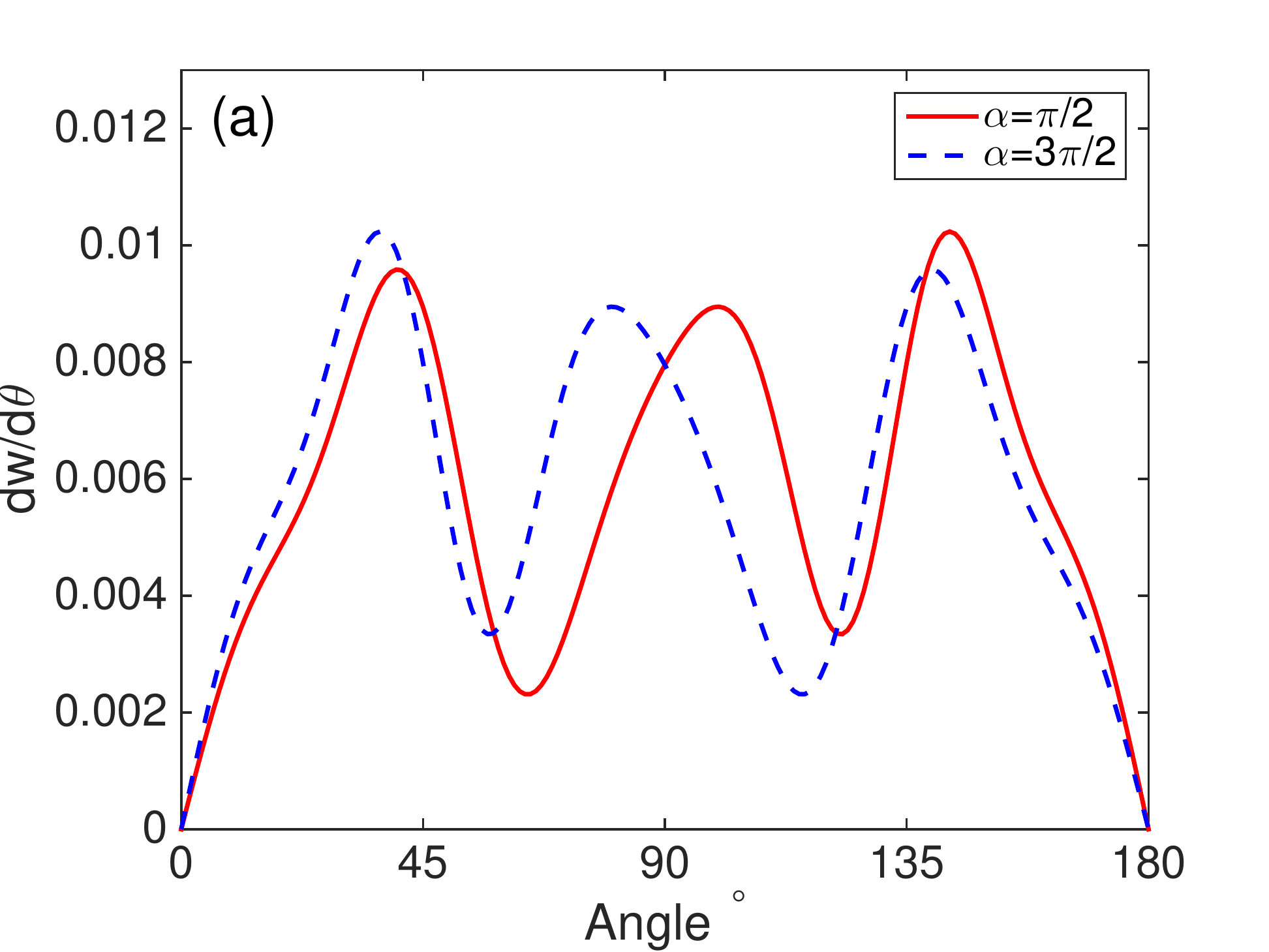}}
\subfigure{\includegraphics[width=2.45in]{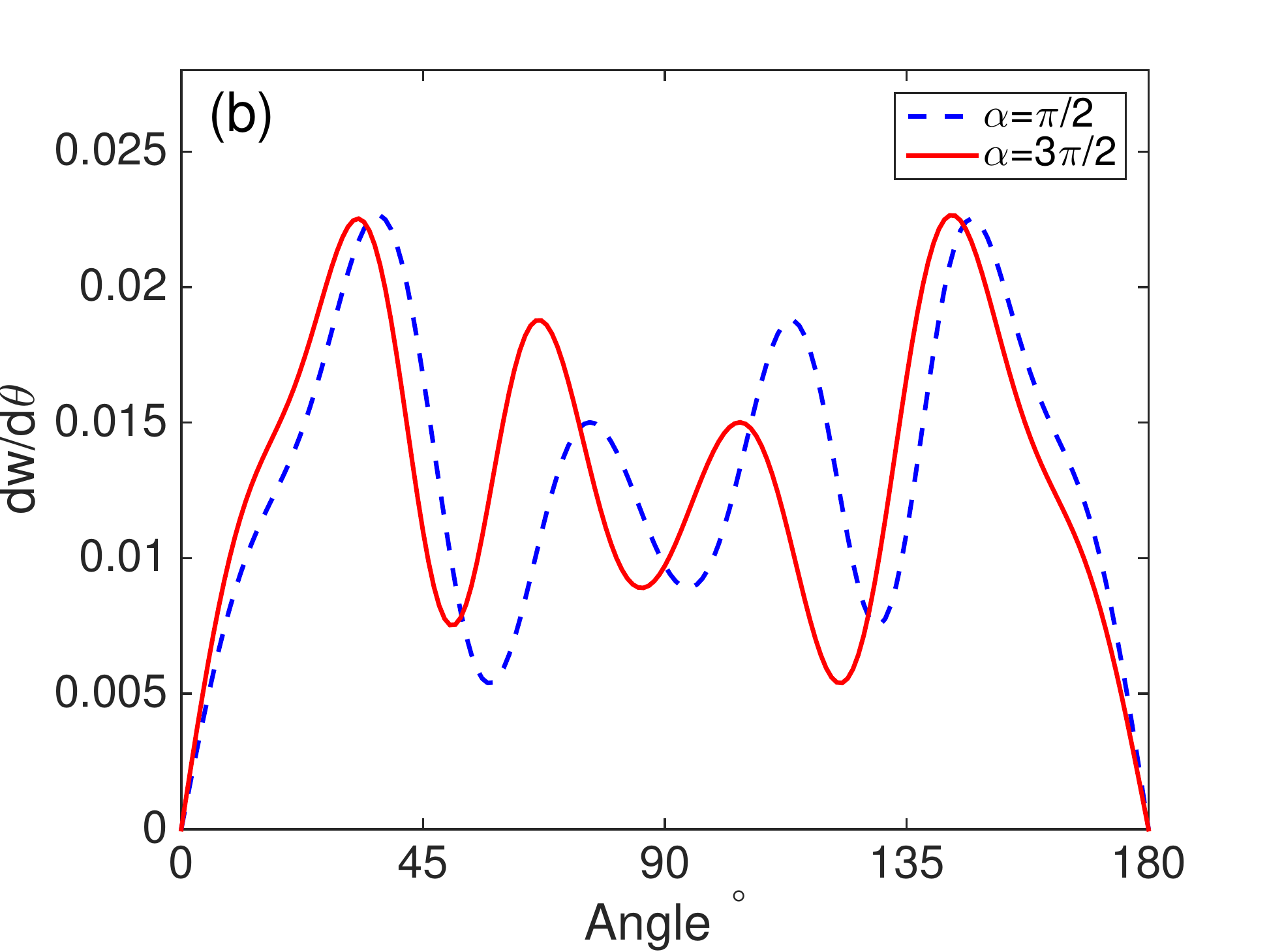}}
\subfigure{\includegraphics[width=2.45in]{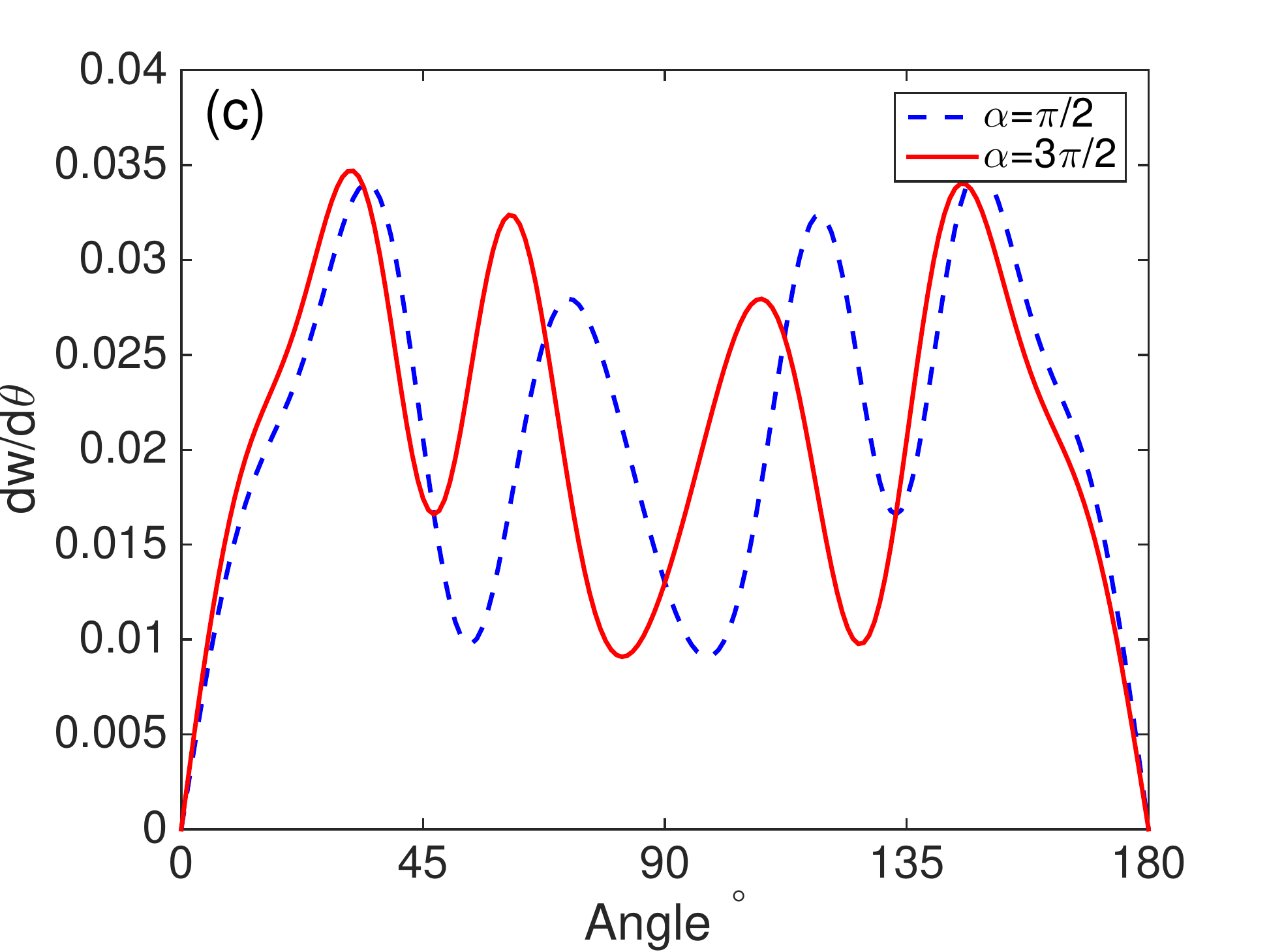}}}
\hspace*{-0.5in}
\mbox{\subfigure{\includegraphics[width=2.45in]{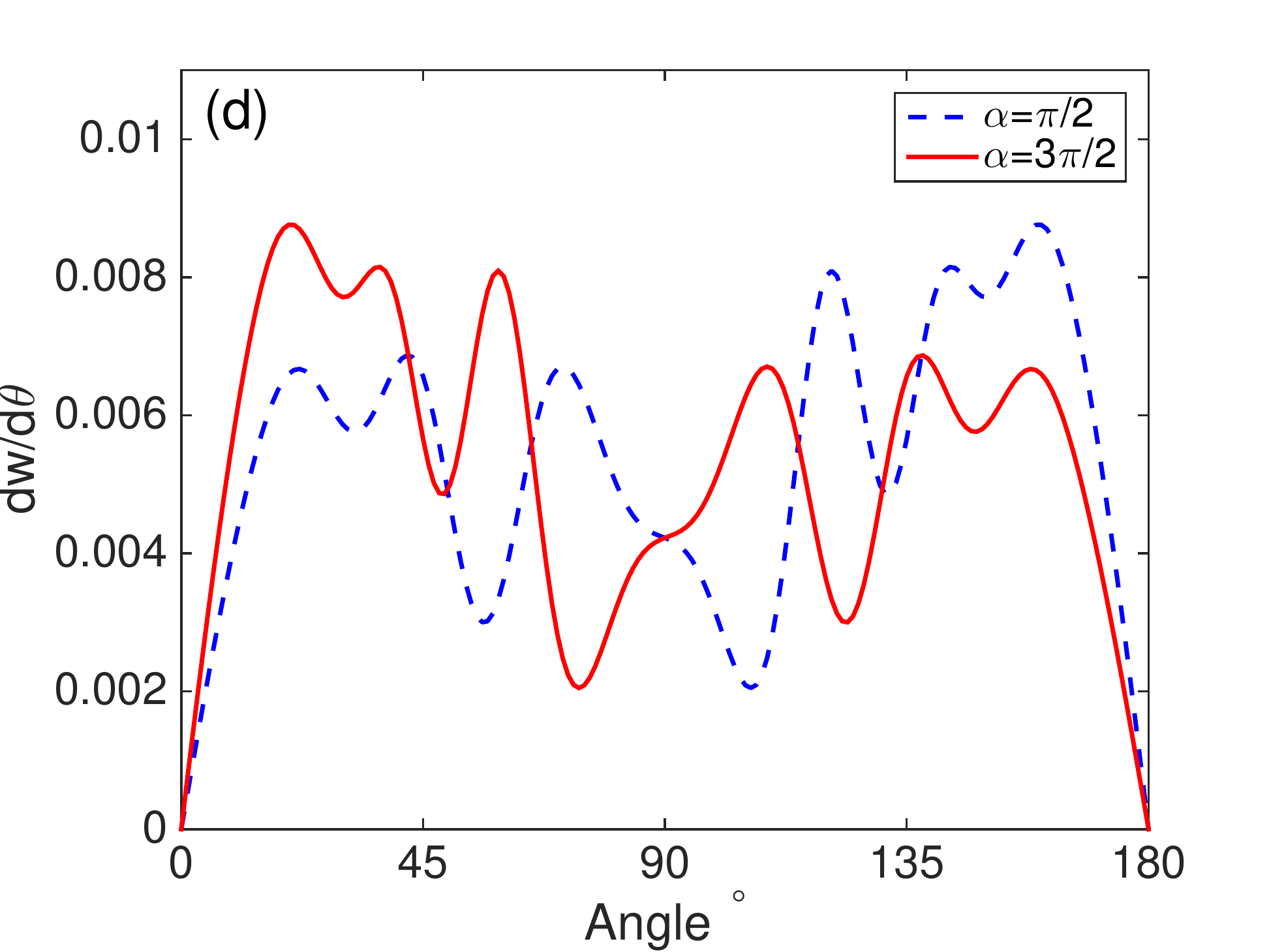}}
\subfigure{\includegraphics[width=2.45in]{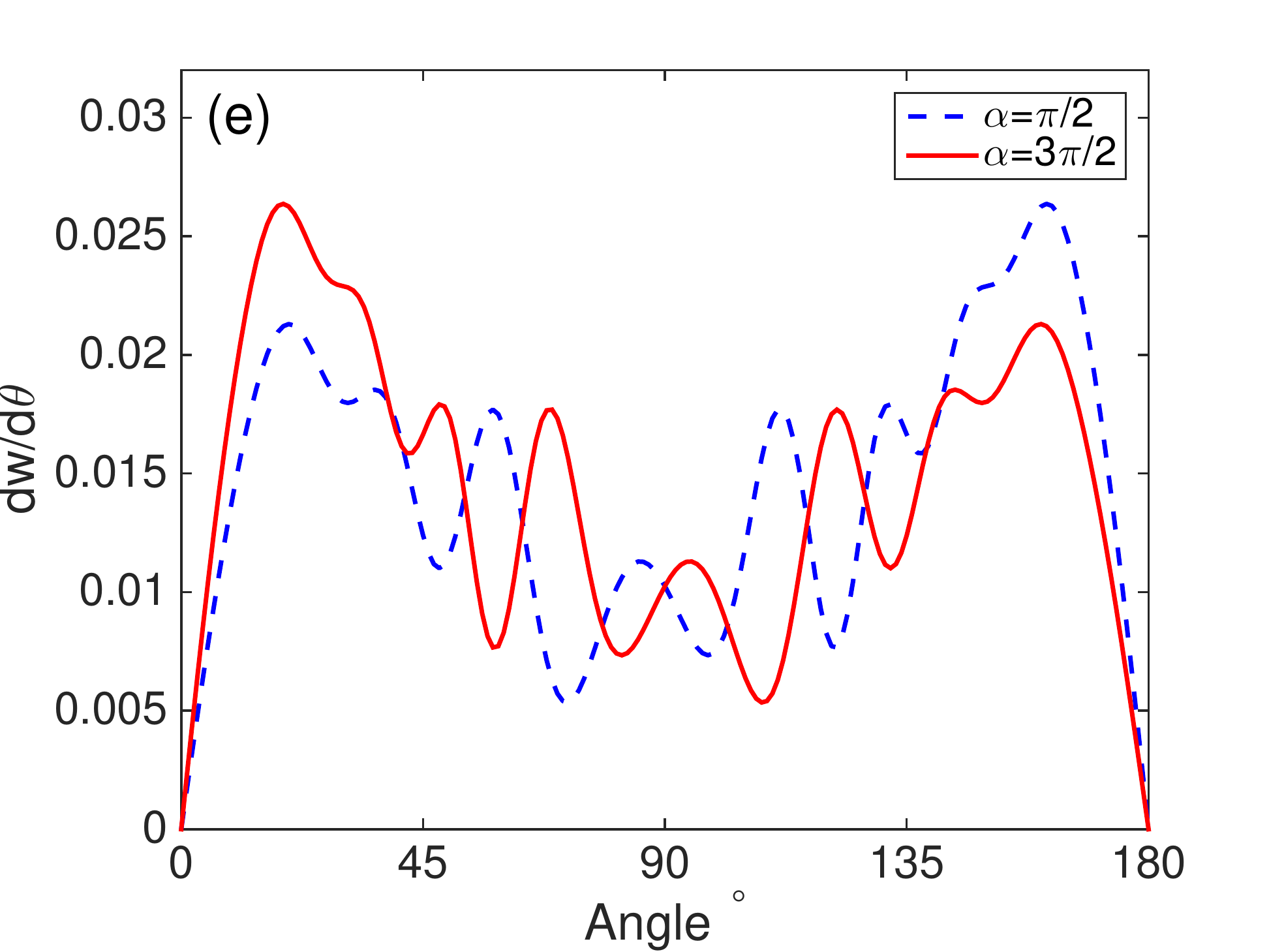}}
\subfigure{\includegraphics[width=2.45in]{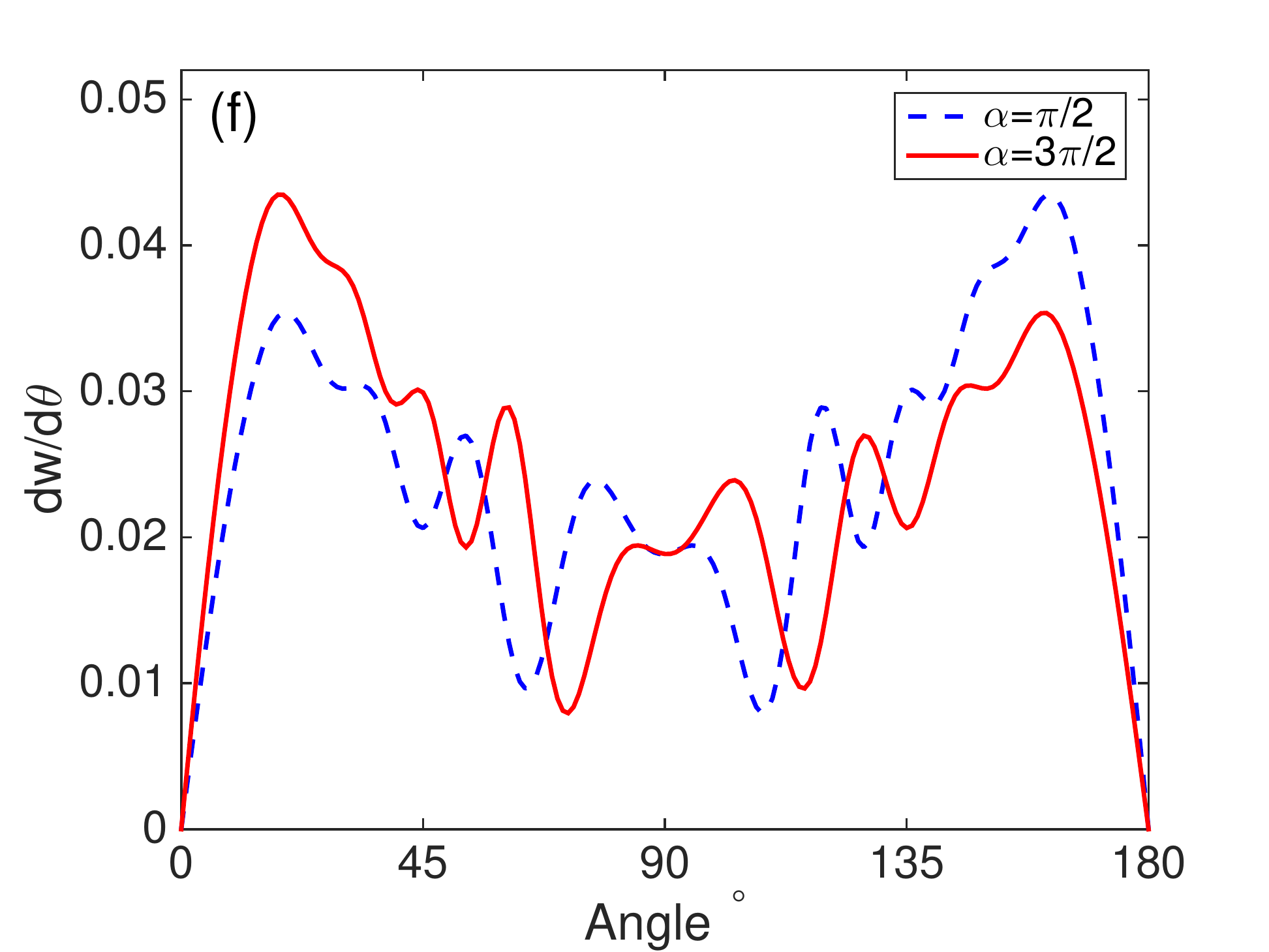}}}
\caption{(Color online) Angular distributions as in Fig.~\ref{fig:angle} but
for $\alpha =\pi/2$ (dashed) and $3\pi/2$ (solid).}
\label{fig:anglecep}
\end{figure}

Figure~\ref{fig:anglecep} displays the angular distributions computed for CEP
values $\alpha =\pi /2$ and $3\pi/2$, and corrects Fig.~6 or
Ref.~\cite{shearer}. While the shapes are very different, the degree of
asymmetry on the angular distributions for these CEP values is consistent
with that seen in Ref.~\cite{shearer}.

\begin{figure}[ht]
\hspace*{-0.5in}
\mbox{\subfigure{\includegraphics[width=2.45in]{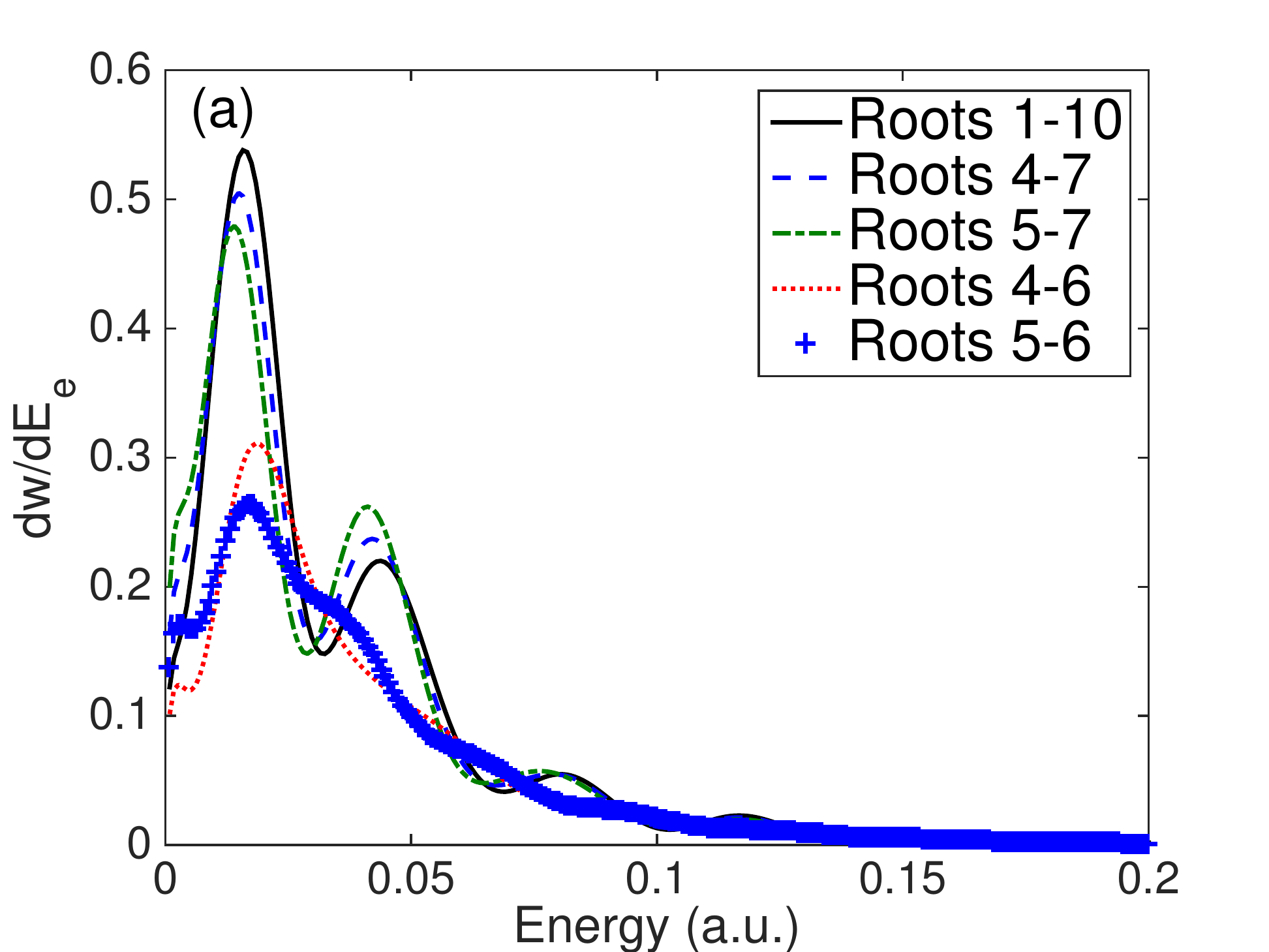}}
\subfigure{\includegraphics[width=2.45in]{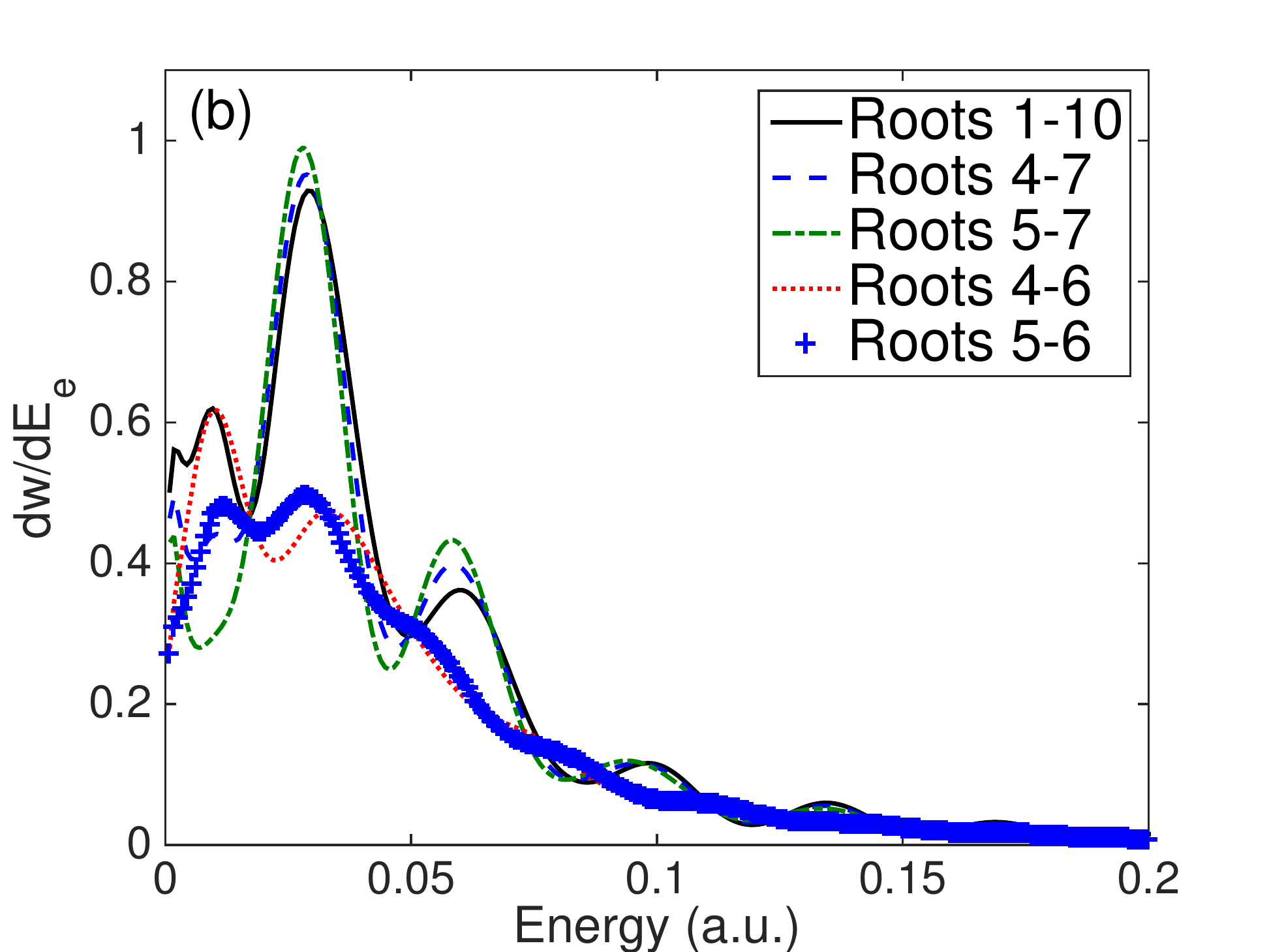}}
\subfigure{\includegraphics[width=2.45in]{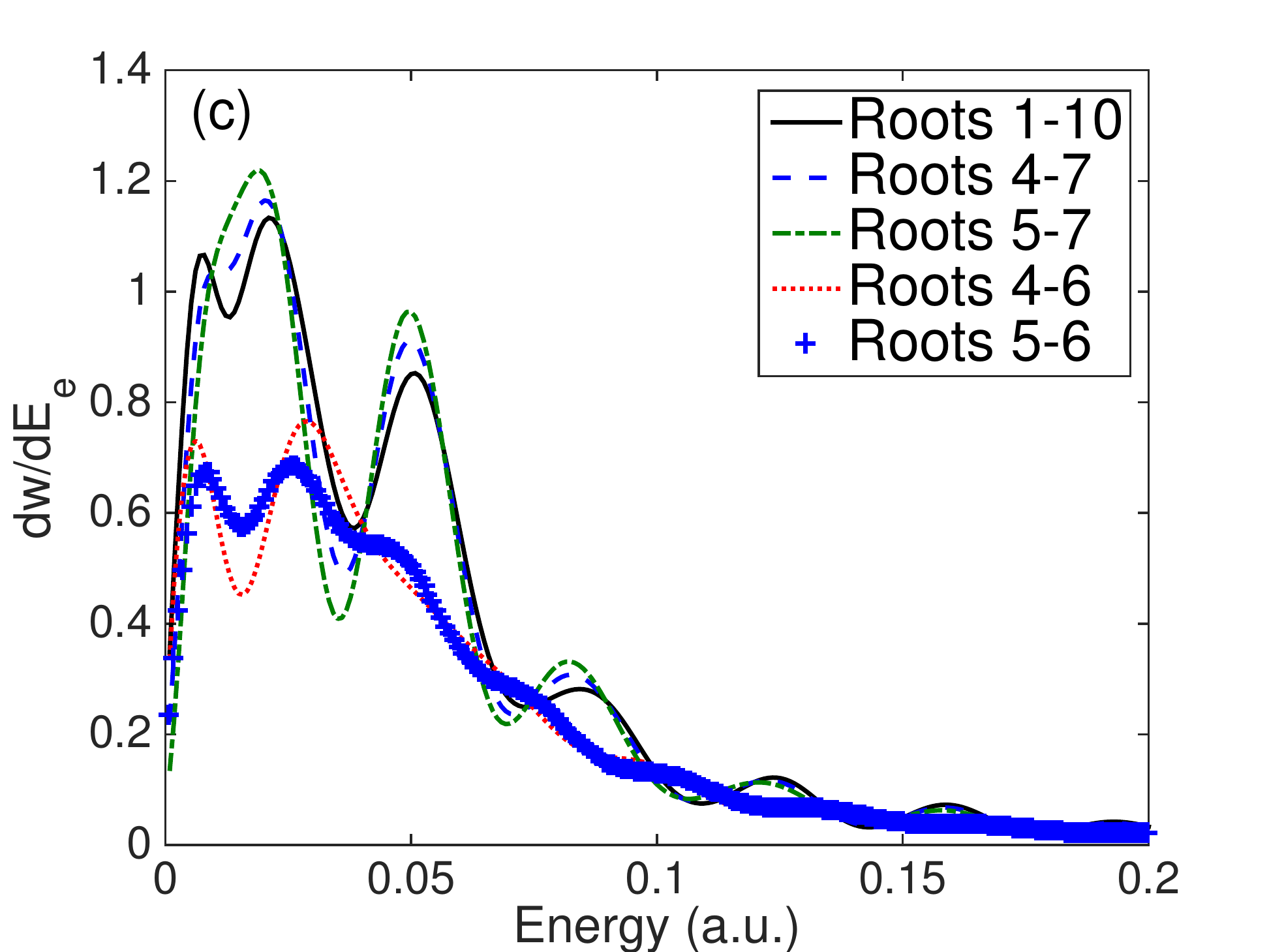}}}
\hspace*{-0.5in}
\mbox{\subfigure{\includegraphics[width=2.45in]{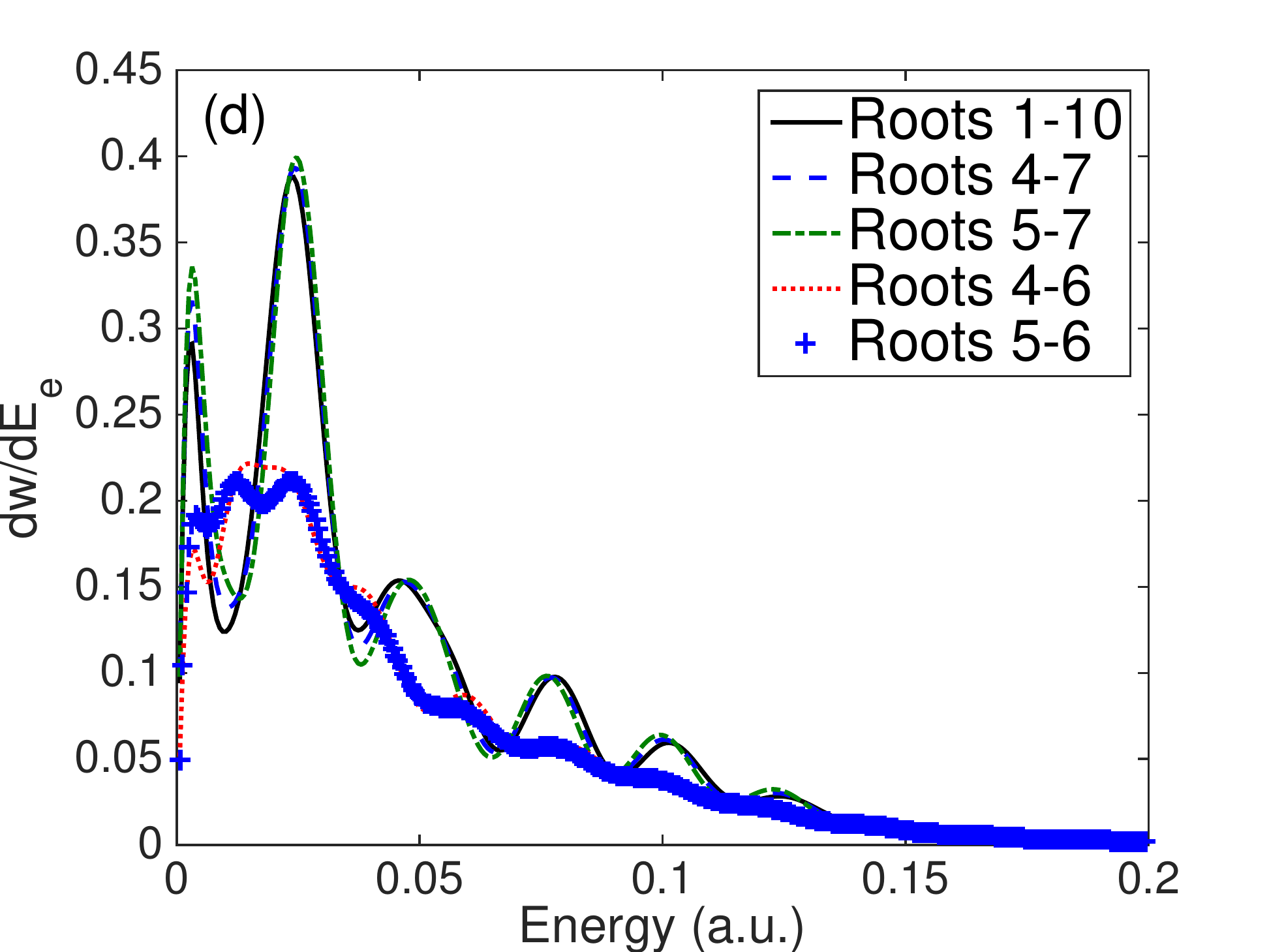}}
\subfigure{\includegraphics[width=2.45in]{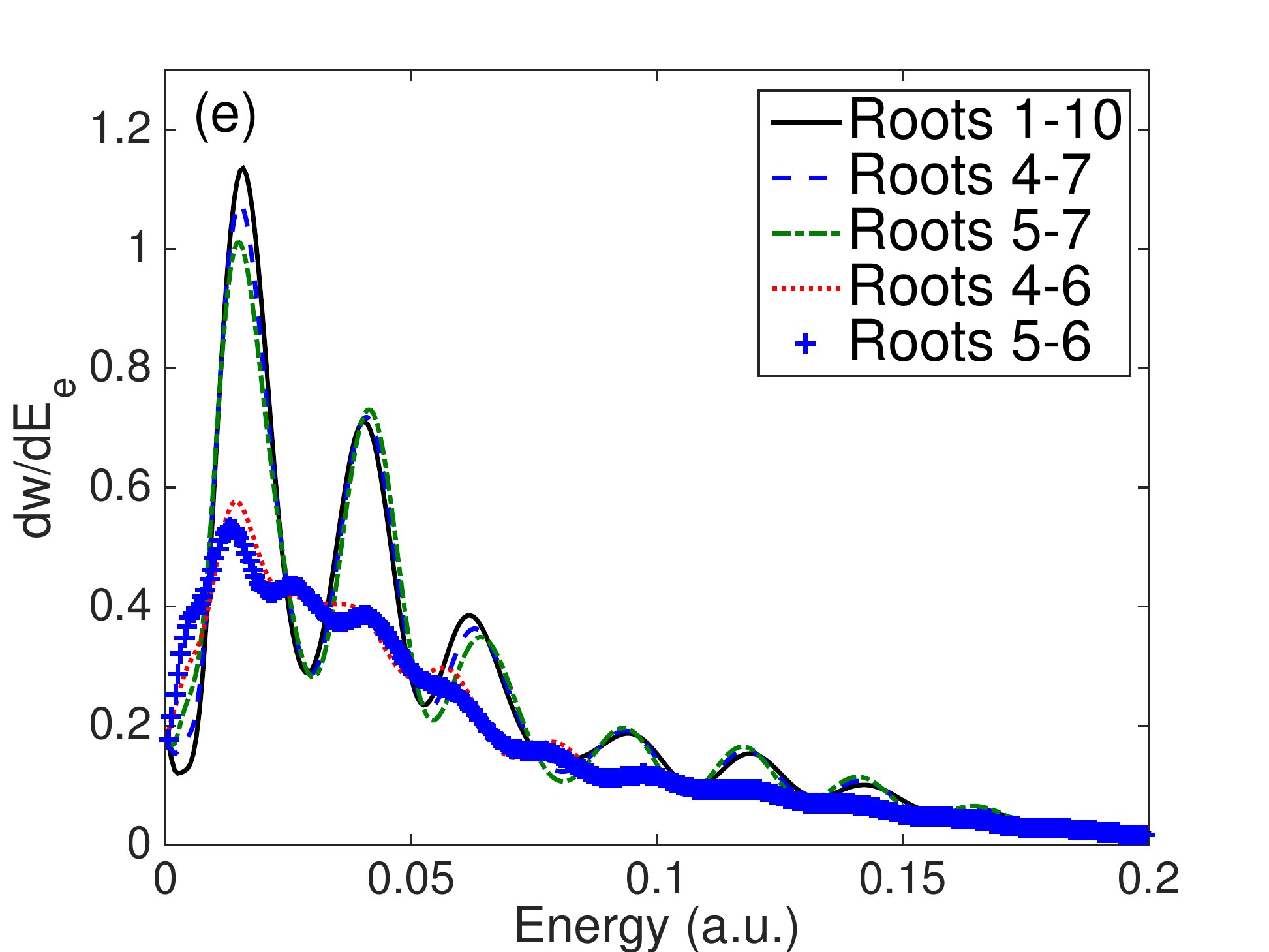}}
\subfigure{\includegraphics[width=2.45in]{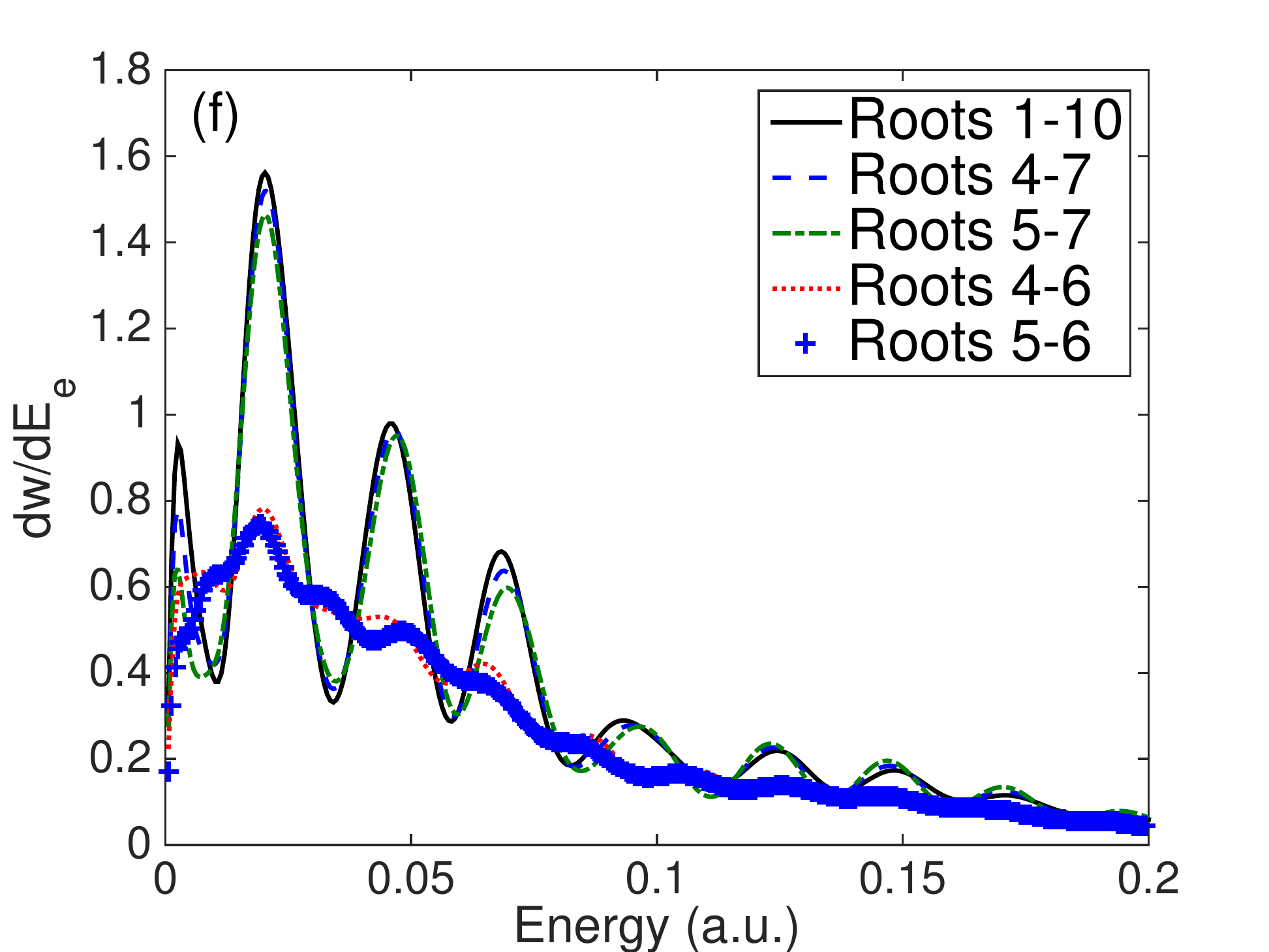}}}
\caption{(Color online) Photoelectron energy spectra of F$^-$ for the same
laser pulse parameters for each panel as in Fig. \ref{fig:angle}.
Partial contributions from selected saddle points (``roots'') are also shown.}
\label{fig:energy}
\end{figure}

Figure~\ref{fig:energy} presents the photoelectron energy spectra for a
four-cycle pulse with $\alpha=0$ and the same wavelengths and intensities as
in Fig.~\ref{fig:angle}. It also shows the spectra obtained by
including only 2, 3 or 4 saddle points closest to the centre of the pulse.
Comparing with Fig.~7 of Ref.~ \cite{shearer}, we see that the shapes
and magnitudes of the correct spectra are quite different from those reported
in Ref.~\cite{shearer}. We note that the spectra shown in
Fig.~\ref{fig:energy} (c) and (f) (corresponding to intensity
$1.3\times 10^{13}$~W/cm$^2$ and wavelength 1300 and 1800~nm, respectively)
show better agreement with those calculated using RMT \cite{ola} in the
low-energy region (see Fig.~2(c) and (d) of \cite{erratum}).

\begin{center}
\begin{table}[t!]
\caption{Total detachment probabilities for photodetachment of F$^-$ calculated
for a four-cycle pulse at wavelengths 1300 and 1800~nm, and different
peak intensities $I$ and CEP phases $\alpha=0$ and $\pi/2$. The last column
shows the detachment probabilities per period for a long pulse.}
\label{tab:totalprobf}
\begin{ruledtabular}
\begin{tabular}{ccccccc}
$\lambda $ & $I$  & $n_{\rm min}$ & $w$ & $w$ &
$w_\text{no-int}$\footnote{Obtained by adding modulus-squared contributions
from each saddle point in Eq.~(\ref{eq:glebshort}).} & $dw/dt(2\pi/\omega )$ \\
(nm) & (W/cm$^2$) & & ($\alpha=0$) & ($\alpha=\pi/2$) & \\
\hline
1300 & 7.7 $\times$ 10$^{12}$ & 5 & 0.0178 & 0.0185 & 0.0174 & 0.0181 \\
     & 1.1 $\times$ 10$^{13}$ & 6 & 0.0443 & 0.0412 & 0.0448 & 0.0416 \\
     & 1.3 $\times$ 10$^{13}$ & 6 & 0.0687 & 0.0659 & 0.0683 & 0.0753 \\
 \hline
1800 & 7.7 $\times$ 10$^{12}$ & 9 & 0.0159 & 0.0170 & 0.0165 & 0.0165\\
     & 1.1 $\times$ 10$^{13}$ & 10 & 0.0468 & 0.0461 & 0.0475 & 0.0497\\
     & 1.3 $\times$ 10$^{13}$ & 11 & 0.0752 & 0.0768 & 0.0754 & 0.0789\\
\end{tabular}
\end{ruledtabular}
\end{table}
\end{center}

For completeness, Table~\ref{tab:totalprobf} gives the total detachment
probability for F$^-$ for all wavelengths and intensities considered, with CEP
values $\alpha$=0 and $\pi/2$ ($\alpha =3\pi/2$ and $\pi/2$ give equivalent
results).
As shown above, correct phases
of the terms in the amplitude are crucial for the shapes of the photoelectron
momentum, angular and energy distributions. However, they play a relatively
small role in the total detachment probability. We have checked that if
the latter is calculated by omitting the $(\pm)^{l+m}$ factor in
Eq.~(\ref{eq:glebshort}), the total detachment probability is within 1-2\% of
the values given in Table~\ref{tab:totalprobf}. In fact, the total detachment
probability obtained by neglecting the interference terms [i.e., by
adding the modulus-squared values of the individual saddle-point contributions
in Eq.~(\ref{eq:glebshort})], $w_\text{no-int}$, is within few per cent of the
correct value. This shows that the interference of the photoelectron wave
packets produced at different laser-field maxima does not lead to much
suppression of enhancement of electron emission, but only to some spatial
redistribution of the photoelectron flux. The values of $w_\text{no-int}$
shown in the second last column of Table~\ref{tab:totalprobf} are also
practically independent of the CEP (with differences $\sim 0.01\%$). 

Shown in the last column of Table~\ref{tab:totalprobf} are
the detachment probabilities \textit{per period} determined from the KTA
detachment rates in a long periodic pulse, $dw/dt$ \cite{gleb}. They are
close to the detachment probabilities in the four-cycle pulse, which implies
that effectively, the detachment in the four-cycle pulse is dominated by
the central, strongest-field cycle. A similar agreement was seen in
Ref.~\cite{shearer1} (Table II) which compared detachment probabilities of
H$^-$ in a five-cycle pulse with the corresponding one-period long-pulse
probabilities from Ref.~\cite{gleb}.

Additionally, it is interesting to note that the total detachment probabilities
in the short pulse are slightly greater for $\alpha=\pi/2$ than for $\alpha=0$
if $n_{\rm min}$ is odd, but slightly smaller when $n_{\rm min}$ is even.
This effect is entirely due to intereference. It can be explained by the fact
that for $\alpha=\pi/2$ the time-dependent electric field
${\bf E}(t)=-d{\bf A}/dt$ acquires its maximum peak magnitude twice within two
central half-cycles, whereas for $\alpha=0$ the field reaches its peak value
once at the middle of the pulse. By comparing the angular distributions
(Figs.~\ref{fig:angle} and \ref{fig:anglecep}) for odd and even $n_{\rm min}$
we see that for $\alpha=\pi/2$, constructive interference is more prominent for
odd $n_{\rm min}$ near the $\theta=\pi/2$  direction relative to the field, while
for even $n_{\rm min}$, destructive interference is more pronounced for angles
near $\theta=\pi/2$. This gives slightly higher (lower) total detachment
probabilities seen in Table~\ref{tab:totalprobf} for $\alpha=\pi/2$ when
$n_{\rm min}$ is odd (even).

In conclusion, the photoelectron spectra presented in Ref.~\cite{shearer}
were incorrect, in part due to the omission of the $m$-dependent phase
factor in the sum over the saddle points that gives the amplitude. Using the
correct phase factor is crucial for obtaining correct interference features of
the photoelectron momentum and angular distributions.

\section*{Acknowledgments}
We thank H. W. van der Hart and A. C. Brown for useful discussions.
S. M. K. Law acknowledges funding from a DEL-NI studentship.

\end{document}